\documentclass[12pt]{article}
\usepackage[utf8]{inputenc}
\usepackage{amsmath}
\usepackage{amssymb}
\usepackage{graphicx}
\usepackage{comment}
\usepackage{booktabs}
\usepackage{here}

\newcommand{\del}{\partial}

\makeatletter

\DeclareTextSymbolDefault{\textquotedbl}{T1}

\numberwithin{equation}{section}

\@ifundefined{date}{}{\date{}}

\usepackage{braket}
\usepackage{bm}
\usepackage{bbm}

\usepackage{epsfig}
\usepackage{pstricks}
\usepackage{color}
\usepackage{tikz}

\usepackage{hyperref}

\newcommand {\nn} {\nonumber}

\newcommand{\aln}[1]{\begin{align}#1\end{align}}

\@ifundefined{showcaptionsetup}{}{%
 \PassOptionsToPackage{caption=false}{subfig}}
\usepackage{subfig}
\makeatother

\begin{document}
\begin{titlepage}
\renewcommand{\thefootnote}{\fnsymbol{footnote}}

\begin{flushright} 
  KEK-TH-2618
\end{flushright} 

\vspace{0.5cm}

\begin{center}
  {\bf \large Preconditioned flow as a solution to the hierarchical growth problem
 in the generalized Lefschetz thimble method}
\end{center}

\vspace{0.5cm}


\begin{center}
         Jun N{\sc ishimura}$^{1,2)}$\footnote
          { E-mail address : jnishi@post.kek.jp},
         Katsuta S{\sc akai}$^{1,3)}$\footnote
          { E-mail address : sakai.las@tmd.ac.jp} and
         Atis Y{\sc osprakob}$^{2,4)}$\footnote
          { E-mail address : ayosp@phys.sc.niigata-u.ac.jp}


\vspace{1cm}

$^{1)}$\textit{KEK Theory Center,
Institute of Particle and Nuclear Studies,}\\
{\it High Energy Accelerator Research Organization,\\
  1-1 Oho, Tsukuba, Ibaraki 305-0801, Japan}

$^{2)}$\textit{Graduate Institute for Advanced Studies, SOKENDAI,\\
1-1 Oho, Tsukuba, Ibaraki 305-0801, Japan}

$^{3)}$\textit{College of Liberal Arts and Sciences, 
Tokyo Medical and Dental University,}\\
{\it 2-8-30 Kounodai, Ichikawa, Chiba 272-0827, Japan}

$^{4)}$\textit{Department of Physics, Niigata University,}\\
{\it
8050 Igarashi 2-no-cho, Nishi-ku, Niigata-shi, Niigata 950-2181, Japan}
\end{center}

\vspace{0.5cm}

\begin{abstract}
  \noindent
The generalized Lefschetz thimble method is a promising approach that
attempts to solve the sign problem
in Monte Carlo methods
by deforming the integration contour
using the flow equation.
Here we point out 
a general problem that occurs
due to the property of the flow equation,
which extends a region on the original contour
exponentially to a region on the deformed contour. 
Since the growth rate for each eigenmode is governed 
by the singular values of the Hessian of the action,
a huge hierarchy in the singular value spectrum,
which typically appears for large systems,
leads to various technical problems in numerical simulations.
We solve this hierarchical growth problem by preconditioning the flow
so that the growth rate becomes identical for every eigenmode.
As an example, we show that the preconditioned flow enables us
to investigate the real-time quantum evolution of an 
anharmonic oscillator with the system size 
that can hardly be achieved by using the original flow.
 

\end{abstract}
\vfill
\end{titlepage}
\vfil\eject


\setcounter{footnote}{0}

\section{Introduction}

In theoretical physics, we are often faced with
a system which cannot be solved analytically.
In that case, we have to either perform perturbative calculations
by taking some limits in the parameter space, 
or make some plausible simplifications so that
the system becomes analytically tractable.
Such approximations, however,
quite often obscure the physics we are most interested in.
In this regard,
numerical simulation is of particular importance
since it provides us with a powerful tool to investigate various systems
nonperturbatively from first principles without such approximations.

Here we are concerned with numerical simulation of multi-variable integrals 
with some weight, where the number of variables $N$
representing the system size is large. 
In the case with a positive-definite weight
such as the Boltzmann weight that
appears
in classical statistical mechanics,
the numerical integration
can be realized by
Monte Carlo (MC) simulation
based on the importance sampling,
where we interpret the weight as the probability distribution.
However, we often encounter
a weight $w=e^{-S}$ with a complex action $S$, 
which cannot be interpreted as the probability distribution.
A naive prescription is the so-called reweighting,
where we interpret the absolute value of the weight as the probability
distribution,
and take into account the phase factor when we take the average of the observables.
This does not really work for large systems since the phase factor
oscillates violently depending on the generated configurations, which
leads to huge cancellations among them.
Thus,
in order to obtain the correct expectation value,
one needs a huge number of configurations,
which typically grows exponentially with the system size $N$. 
This is the notorious sign problem, which has been hindering
nonperturbative studies of various important systems,
such as finite density QCD, theories with a $\theta$-term,
strongly correlated electron systems,
and the Lorentzian path integrals for the real-time quantum evolution.

In the last decade, various new approaches have been developed to overcome 
the sign problem.
In particular, within the framework of MC methods,
two promising approaches,
the complex Langevin method \cite{Klauder:1983sp, Parisi:1984cs}
and the Lefschetz thimble
method \cite{Witten:2010cx,Cristoforetti:2012su,Cristoforetti:2013wha,Fujii:2013sra}
have been studied intensively\footnote{
As a promising approach without MC simulation, 
the tensor renormalization group has been developed \cite{Levin:2006jai, PhysRevB.86.045139, PhysRevLett.115.180405, Adachi:2019paf, Kadoh:2019kqk}. 
The basic idea is to rewrite the integral
as a network of tensors and to perform a 
coarse-graining procedure iteratively using the singular value decomposition.
Since there is no need to interpret the weight as the probability
distribution,
the method is free from the sign problem from the outset.}.
A common feature of these two approaches is that
one complexifies 
the variables and extends the weight and the observables
as holomorphic functions of the complexified variables. 
In the complex Langevin method,
one generates complexified
configurations using the Langevin equation
with the drift term given by the gradient of
the action.
While the cost of this method is O($N$) as in typical MC simulation,
it is applicable only to 
a limited class of systems due to
the conditions for
justification \cite{Aarts:2009uq,Aarts:2011ax,Nagata:2015uga,Nagata:2016vkn,Scherzer:2018hid,Scherzer:2019lrh,Seiler:2023kes}. 

On the other hand, the Lefschetz thimble method
is based on the Picard-Lefschetz theory,
in which one deforms the integration contour in the complex space
to a set of steepest descent paths
emanating from some saddle points of the action.
These paths are called the Lefschetz thimbles (or thimbles in short).
The phase of the complex weight is constant along each thimble,
and the sign problem is solved
as far as the phase factor coming from the
integration measure can be taken into account by reweighting \cite{Fujii:2013sra}. 

In practice, the
contour deformation
can be realized
by solving the so-called anti-holomorphic gradient flow equation 
for the variables \cite{Alexandru:2015sua},
where the amount of flow (the flow time)
plays the role of the deformation parameter.
While the Lefschetz thimbles are obtained in the long flow time limit,
one can also use a deformed contour obtained at
finite flow time of the order of $\log N$,
which is expected to be long enough to solve the sign problem
for the system size $N$.
This is called the generalized thimble method (GTM),
which has a big advantage over the
original Lefschetz thimble method
in that it does not require prior knowledge of the Lefschetz thimbles.
(See Ref.~\cite{Alexandru:2020wrj} for a review.)

When there are more than one thimbles
that contribute to the integral,
the GTM
suffers from
an ergodicity problem
for a large flow time
due to the infinite potential barriers between different thimbles.
For a relatively small system,
one can just choose 
the flow time
long enough to solve the sign problem but not too long
so that the ergodicity problem can be avoided \cite{Alexandru:2015sua}.
%
The range of the flow time that can avoid both problems,
however, shrinks
as the system size increases and eventually vanishes.
%
In order to solve both problems even in that case,
one can integrate
over the flow time\footnote{This proposal is a significant
improvement over
the related ones \cite{Fukuma:2017fjq,Alexandru:2017oyw,Fukuma:2019uot}
based on tempering with respect to the flow time,
which requires the calculation of the Jacobian when one swaps the replicas.},
which amounts to treating the flow time as an extra dynamical variable
in the simulation \cite{Fukuma:2020fez,Fukuma:2021aoo}.
%

As an efficient algorithm for numerical simulations in general,
the Hybrid Monte Carlo (HMC) algorithm is widely used.
In this algorithm, one uses a fictitious Hamilton dynamics
to update the configuration.
When one applies this algorithm to the GTM,
there are actually two approaches depending on whether one considers
the Hamilton dynamics on the deformed contour
or on the original contour.
In the first approach,
one has to solve the Hamilton dynamics of
a constrained system in order to make sure that
the configuration remains on the deformed
contour \cite{Fujii:2013sra,Fukuma:2019uot}.
This requires some complicated calculations since the deformed
contour is given only implicitly by solving the
gradient flow equation.
A big advantage of this approach, however, is that the probability distribution
of the generated configurations
automatically 
includes the modulus of the Jacobian associated with the flow of variables,
which is not the case
in the second approach
as we explain below.

In the second approach,
one can just solve the Hamilton dynamics of
an unconstrained system on the real axis, which makes this part of
the algorithm much simpler than in the first approach.
The force term of the Hamilton equation has to be calculated
by taking the derivative of the action on the deformed contour
with respect to the variables on the real axis, 
which is actually possible without large computational cost
if one uses the idea of backpropagation \cite{Fujisawa:2021hxh}.
A drawback of this second approach, however, is that
the probability distribution of the generated configurations
does not include 
the modulus of the Jacobian associated with the flow of variables,
and therefore it has to be included by reweighting
together with the phase of the Jacobian.
In fact, the modulus may fluctuate considerably
depending on the generated configurations.
In that case, only a few configurations that have a large modulus of the Jacobian
dominate the ensemble, and one cannot increase the statistics efficiently.
This is the so-called overlap problem that
can occur in general when one uses reweighting.
In what follows,
we call these
two approaches the on-thimble approach\footnote{This is a bit of
  abuse of the word ``thimble'' since here it actually represents
  a deformed contour obtained by the
  gradient flow
  at finite $\tau$.}
and the on-axis approach, respectively.

In this paper,
we point out a general problem that occurs when the GTM
is applied to large systems.
In the GTM, the deformation of the integration contour
by the
gradient flow equation is really the key to
solve the sign problem.
The crucial feature of the gradient flow equation
that makes this possible
is that it maps a small region on the real axis
to an exponentially large region on the deformed contour as the flow time
becomes longer.
The growth rate of this exponential behavior, however,
depends on the eigenmode, and it is governed by the singular values of
the Hessian of the action.
As the system size increases, the singular value spectrum typically
exhibits a huge hierarchy.
As a consequence,
if we choose the flow time long enough
to solve the sign problem associated with the eigenmodes
corresponding to small singular values,
the eigenmodes corresponding to large singular values
tend to diverge and easily get out of control\footnote{This
  problem
  should not be confused with
  the blow-up problem of the flow equation
  discussed in Ref.~\cite{Tanizaki:2017yow}, which occurs
  when the integrand
  becomes zero at some point in the complexified configuration space.
  In fact,
  we consider that the blow-up problem does not occur
  if one uses the preconditioned flow
  that we propose in this paper
  as we discuss 
  in Appendix \ref{sec: blow-up}.}.

We show that this hierarchical growth problem of the GTM
can be solved by preconditioning the
gradient flow equation.
For that, we first point out that
we are free to
introduce a Hermitian positive-definite kernel on the
right-hand side of the gradient flow equation without spoiling
its property that is necessary to solve the sign problem.
One can actually use this freedom to make
the growth rate
identical for
every eigenmode,
which implies that
the hierarchical growth problem can be solved by the preconditioned flow.
Moreover, when applied to the on-axis approach,
the preconditioned flow suppresses the fluctuation of the Jacobian,
which can otherwise cause the overlap problem as we mentioned above.
Actual implementation of this preconditioned flow can be done
by using the techniques developed for Rational Hybrid Monte Carlo
algorithm, which is widely used in numerical simulation.
In order to demonstrate
how the preconditioning works,
we apply the GTM
to the real-time quantum evolution of an anharmonic oscillator\footnote{See also
  Refs.~\cite{Alexandru:2016gsd,Alexandru:2017lqr,Mou:2019tck,Mou:2019gyl,Woodward:2022pet}
  for calculations of the real-time quantum evolution using the GTM
  based on the Schwinger-Keldysh formalism, which actually simplifies the
  simulation \cite{Mou:2019tck}.}.
In particular, we show that
the preconditioning enables us to
simulate a large system
that cannot be achieved otherwise.
We also show that
the computational cost
for generating configurations
grows only linearly with the system
size $N$
for a local system if we implement 
the preconditioning in
the flow appropriately.
Preliminary discussions on the precondition flow
are presented in our previous paper \cite{Nishimura:2023dky}
and a proceedings article \cite{Nishimura:2023jtu},
where
we establish a new picture of
quantum tunneling in the real-time path integral using the GTM.


The rest of this paper is organized as follows. 
In Section \ref{sec: subtlety},
we discuss the hierarchical growth problem of the original flow 
due to the hierarchy in the singular value spectrum of the Hessian. 
In particular, we show that the problem becomes severer when the system
size increases using a simple example of a harmonic oscillator.
In Section \ref{sec: precond}, we explain
our proposal of the preconditioned flow equation,
and discuss how it can be implemented in practice.
%
In Section \ref{sec: simulation},
we demonstrate how the preconditioning solves the hierarchical growth problem
of the flow as well as the overlap problem in the on-axis approach
by applying the GTM to
the real-time quantum evolution of an anharmonic oscillator.
%
Section \ref{sec: summary} is devoted to a summary and discussions.
In Appendix \ref{sec: blow-up}, we
discuss the absence of the so-called
blow-up problem
in
the preconditioned flow.
In Appendix \ref{sec: thimbles}, we show that
the preconditioned flow actually changes the thimble
and yet it solves the sign problem using a simple example. 
In Appendix \ref{sec: HMC-on-axis-review}, we provide a brief review
on the application of the HMC to the GTM based on the on-axis approach.
In Appendix \ref{sec: forces},
we discuss how the idea of backpropagation for calculating the force
term in the HMC algorithm works with the preconditioned flow.
In Appendix \ref{sec: parameters}, we present the
parameters
of the GTM
used in our simulations.


\section{The hierarchical growth problem in the GTM}
\label{sec: subtlety}

In this section, we discuss the crucial properties
of the
gradient flow equation
that make it possible to solve the sign problem.
In particular, we show
that it maps a region on the real axis
to a region on the deformed contour,
which becomes exponentially large with the flow time.
Moreover the growth rate for each eigenmode is governed
by the singular value spectrum of the Hessian of the action.
When
the system size $N$ is large,
the growth rate tends to exhibit a large hierarchy,
which is
the hierarchical growth problem.

\subsection{Hierarchical growth property of the flow equation}
\label{sec:hierarchical-prop}

Let us consider a partition function
with a complex action $S(x)\in\mathbb{C}$ defined by
\aln{
  Z=\int_{\mathbb{R}^N}d^Nx\,e^{-S(x)} \ .
  \label{eq:general-path-integral}
}
In the GTM,
one deforms the integration contour by using
the
gradient flow equation
\aln{
  \frac{d z_j(\sigma)}{d \sigma}
  =\overline{\frac{\partial S(z(\sigma))}{\partial z_j}} \ .
\label{eq: standard-flow}
}
By solving this equation
from $\sigma=0$ to $\sigma=\tau$
with the initial condition $z_j(0)=x_j$,
we obtain a map $x \mapsto z(x,\tau) \equiv z(\tau)$.
Then the deformed contour is defined by
$\Sigma_\tau  = \{  z(x,\tau) | x \in \mathbb{R}^N  \} $,
where $\tau$ is the so-called flow time.
Due to Cauchy's theorem, the partition function
can be rewritten as
\aln{
  Z&=\int_{\Sigma_\tau}d^Nz\,e^{-S(z)}  \ .
  \label{part-fn-deformed}
}

The crucial
property of the flow equation is
\aln{
  \frac{d S(z(\sigma))}{d \sigma}
  = \frac{\partial S(z(\sigma))}{\partial z_j}
  \frac{ d z_j(\sigma)}{d \sigma}
  =   \left|\frac{\partial S(z(\sigma))}{\partial z_j}\right|^2
  > 0 \ ,
\label{eq: flow-essence}
} 
which implies that
$\mathrm{Re}\,S(z(\sigma))$ increases monotonically
with increasing $\sigma$,
while $\mathrm{Im}\,S(z(\sigma))$ remains constant.
In the infinite flow time limit $\tau=\infty$,
the deformed contour $\Sigma_{\infty}$
is composed of
a set of Lefschetz thimbles.
Each thimble can be obtained by the flow from some saddle
point (defined by $\frac{\del S(z)}{\del z_j} = 0$),
and hence $\mathrm{Re}\,S(z(\sigma))$ increases monotonically
keeping $\mathrm{Im}\,S(z)$ constant as one flows away from the
saddle point.
This implies that the sign problem is solved on each thimble
as far as
the phase factor coming from the integration measure $d^N z$
in \eqref{part-fn-deformed}
can be treated by reweighting \cite{Fujii:2013sra}. 
The key mechanism for solving the sign problem here
is that there is some point $P$ on $\mathbb{R}^N$ which flows
into a saddle point in the $\tau \rightarrow \infty $ limit,
and
the thimble associated with that saddle point
is obtained by mapping an
infinitesimal vicinity of the point $P$ using the flow.
This property of the flow, together with the fact that
$\mathrm{Im}\,S(z)$ is kept constant along the flow,
makes it possible to solve the sign problem.
In fact, the sign problem can be solve even at finite $\tau$
as far as $\tau$
is large enough to suppress the fluctuation of $\mathrm{Im}\,S(z)$.


Here we discuss in more detail
how a small region on the original contour $\mathbb{R}^N$
is extended
in the deformed contour.
For that, let us consider two points $x$ and $x+\delta x$
on the real axes, which differ only by some infinitesimal displacement $\delta x$.
The displacement
$\zeta(\sigma) \equiv z(x+\delta x,\sigma)-z(x,\sigma)$
of the two points after some flow time $\sigma$
is given by the equation
\aln{
  \frac{d \zeta_j(\sigma)}{d \sigma}
  =\overline { H_{jk}(z(\sigma)) \, \zeta_k(\sigma) }
\label{eq: flow-of-displacement}
}
with $\zeta_j(0)=\delta x_j$, 
where $H$ is the Hessian of the action defined by
\begin{align}
  H_{ij}(z)
  &= \frac{\partial^2 S(z)}{\partial z_i\partial z_j} \ .
  \label{eq:def-Hessian}
\end{align}
In order to solve \eqref{eq: flow-of-displacement},
let us rewrite it as
\begin{align}
  \frac{d}{d \sigma}
\begin{pmatrix}
    \zeta(\sigma)  \\
   \bar{\zeta} (\sigma)  
\end{pmatrix}
&=
{\cal H}(\sigma)
\begin{pmatrix}
   \zeta(\sigma)    \\
    \bar{\zeta}(\sigma)   
\end{pmatrix} \ ,
\label{hgeJ-matrix}
\end{align}
where
we use the vector notation
$\zeta = (\zeta_1 , \cdots , \zeta_N)^\top$,
$\bar{\zeta} = (\bar{\zeta}_1 , \cdots , \bar{\zeta}_N)^\top$
and define the $2N \times 2N$ Hermitian\footnote{Note that the Hessian
  is symmetric $H^\top = H$ and hence $\bar{H}=H^\dagger$.}
matrix ${\cal H}(\sigma)$ as
\begin{align}
{\cal H}(\sigma) &= 
  \begin{pmatrix}
    &
    \overline{H(z(\sigma))}  \\
        H(z(\sigma))  &   
  \end{pmatrix} \ .
\label{def-cal-H}
\end{align}
Thus the solution to
the differential equation \eqref{eq: flow-of-displacement}
can be written down formally as
\begin{align}
\begin{pmatrix}
 \zeta(\tau)  \\
 \bar{\zeta} (\tau)  
\end{pmatrix}
&=
{\cal P} \exp \left( \int_0 ^\tau  d \sigma \, {\cal H}(\sigma) \right)
\begin{pmatrix}
   \delta x \\
   \delta x
\end{pmatrix} \ ,
\label{zeta-formal-sol}
\end{align}
where
$\delta x = (\delta x_1 , \cdots , \delta x_N)^\top$
and ${\cal P} \exp$ represents the path-ordered exponential,
which ensures that ${\cal H}(\sigma)$ with smaller $\sigma$
comes on the right after Taylor expansion.

In order to understand the behavior of the solution \eqref{zeta-formal-sol},
let us diagonalize the Hermitian matrix ${\cal H}(\sigma)$.
For that, we
consider the singular value decomposition (SVD)
of the Hessian $H_{ij}(z(\sigma))$ given as\footnote{This is known
  as the Takagi decomposition,
  which is the SVD for a complex symmetric matrix.}
\begin{align}
  H(z(\sigma)) = U^\top (\sigma) \, \Lambda(\sigma) \, U(\sigma) \ ,
\label{eq:H-SVD}
\end{align}
where $U$ is a unitary matrix and
$\Lambda={\rm diag}(\lambda_1 , \cdots , \lambda_N)$
is a real diagonal matrix with positive\footnote{Here we assume that
  there are no zero singular values for simplicity. However, our idea can
  be easily extended to the case in which this assumption does not hold.
See footnote \ref{footnote:zero-SV}.
}
entries $\lambda_j > 0$.
Using $\Lambda(\sigma)$ and $U(\sigma)$, we can diagonalize ${\cal H}(\sigma)$ as
\begin{align}
  {\cal H}(\sigma)
  &= {\cal U}^\dagger(\sigma)
  \begin{pmatrix}
    \Lambda(\sigma) &    \\
         &     - \Lambda(\sigma)
  \end{pmatrix}  {\cal U}(\sigma) \ , 
  \label{diagonalize-cal-H}  \\
  {\cal U}(\sigma)
  &=
  \frac{1}{\sqrt{2}}
  \begin{pmatrix}
   {\bf 1}_N  & {\bf 1}_N  \\
    - {\bf 1}_N & {\bf 1}_N
  \end{pmatrix} 
  \begin{pmatrix}
    U(\sigma) &   \\
         &   \overline{U}(\sigma)
  \end{pmatrix} \ .
\label{def-cal-U}
\end{align}
This implies that
the displacement grows exponentially with $\sigma$
for a small region of $\sigma$, in which $H_{jk}(z(\sigma))$ can be regarded
as constant.
Moreover,
the growth rate
is governed by the 
singular values of the Hessian $H(z)$, and it depends on the
eigenmode $\zeta_{\rm E}^{(i)} = {\rm Re}(U_{ij} \zeta_j) \sim e^{\lambda_i \sigma}$.
When the singular value spectrum has a large hierarchy,
the growth rate becomes very different for different eigenmodes.
Since the flow time $\tau$ is common to all the eigenmodes,
if we choose $\tau$ in such a way that the sign problem is solved
for the eigenmodes $\zeta_{\rm E}^{(i)}$
with small $\lambda_i$,
the eigenmodes $\zeta_{\rm E}^{(i)}$
with large $\lambda_i$ tend to diverge.
 
This problem is severer in the on-axis approach
since one has to update the configuration on the real axis.
In order to avoid the hierarchical growth problem, one has to choose a very small
step size in solving the Hamilton dynamics of the
HMC
algorithm,
which makes the simulation extremely slow.
In the case of the on-thimble approach, one updates the
configuration on the deformed contour directly, but the problem
may occur when one solves the flow equation to
ensure that the updated configuration is still on the deformed contour.
(See Section \ref{sec: summary} for further discussions.)

\subsection{Example of a harmonic oscillator}
\label{sec:hos-example} 

In this section, we show that the hierarchical growth problem discussed in the previous
section indeed occurs in a simple example of a harmonic oscillator.
This example provides a clear understanding of the problem since the
gradient flow equation can be solved explicitly.

The path integral that describes
the real-time quantum evolution of a harmonic oscillator
is given by
\eqref{eq:general-path-integral}
with the action
\aln{
  S&=-i\sum_{j=0}^{N}\epsilon\left\{  \frac{1}{2}
  \left(\frac{x_{j+1}-x_j}{\epsilon}\right)^2
 - \frac{1}{2} \, m^2 \, \frac{x_{j+1}^2+x_j^2}{2}\right\} \ ,
\label{eq: harmonic-Z}
}
where we fix $x_0 = x_{N+1}=0$ for simplicity and
treat $x_1,\cdots,x_N$ as variables.
The path integral \eqref{eq:general-path-integral} then
represents
the transition amplitude from the origin to the origin
in time $T=\epsilon (N+1)$ up to some known normalization factor. 

In this example,
the Hessian $H(z)$ of the action defined by
\eqref{eq:def-Hessian} is a constant matrix,
which does not depend on the complexified configuration $z_j$.
Furthermore,
it is not only symmetric but also pure imaginary.
Therefore, we can diagonalize it as
\begin{align}
  H = -i  \, O^\top  \, \tilde{\Lambda} \, O \ ,
  \label{H-diagonalization}
\end{align}
where $O\in {\rm SO}(N)$
and $\tilde{\Lambda}$ is a real diagonal matrix
$\tilde{\Lambda} = \mathrm{diag}(\tilde{\lambda}_1, \cdots , \tilde{\lambda}_N)$
with
\aln{
  \tilde{\lambda}_j =
  \epsilon\left\{ \frac{4}{\epsilon^2}\sin^2\frac{\pi j}{2(N+1)}-m^2\right\} \ .
  \label{lambda-hos}
}
Note that $\tilde{\lambda}_j$ can be negative in general,
and the singular values\footnote{The singular value decomposition \eqref{eq:H-SVD}
  of $H= U^\top \, \Lambda \, U$ is given by $U=V \, O$,
  where $V$ is a diagonal unitary matrix with the entries given by
  either $e^{\frac{\pi}{4}i}$ or $e^{-\frac{\pi}{4}i}$ depending on
  whether the corresponding eigenvalue $\tilde{\lambda}_j$ is
  negative or positive, respectively.}
are given by $\lambda_j = |\tilde{\lambda}_j|$.


Using the Hessian $H$ and its diagonalized form \eqref{H-diagonalization},
the action can be written as
\begin{align}
  S= \frac{1}{2}x_jH_{jk}x_k
   = -  \frac{i}{2} \sum_{j=1}^N \tilde{\lambda}_j (y_j)^2  \ ,
\label{eq: harmonic-S}
\end{align}
where we have defined the variables $y_j=O_{jk} x_k$.
Thus the integral \eqref{eq:general-path-integral}
is rewritten as
\aln{
  Z&=
  \int_{\mathbb{R}^N}d^Ny~\exp\left(\frac{i}{2} \sum_{j=1}^N \tilde{\lambda}_j \, y_j^2\right) \ .
}
The flow equation for the complexified variables $z_j(y,\tau)$ is given by
\aln{
  \frac{\partial z_j(y,\sigma)}{\partial \sigma}
  &=i \tilde{\lambda}_j\overline{z_j(y,\sigma)} \ ,
\label{eq: flow-of-z}\\
z_j(y,0)&=y_j \ ,
}
whose solution
can be readily obtained as
\aln{
  z_j(y,\sigma)=\left\{\cosh(\tilde{\lambda}_j\sigma)
  + i\sinh(\tilde{\lambda}_j\sigma)\right\} \, y_j \ .
  \label{solution-z-y}
}
Note that the gradient flow magnifies the displacement $\delta y_j$ on the real axis
by a factor of
\begin{align}
  R_j(\sigma) \equiv
  \sqrt{\cosh^2(\tilde{\lambda}_j\sigma) + \sinh^2(\tilde{\lambda}_j\sigma)  }
  = \sqrt{\cosh(2\tilde{\lambda}_j\sigma) } \sim
  e^{\lambda_j \sigma} \ ,
  \label{magnification-factor}
  \end{align} 
which grows exponentially with $\sigma$, and the growth rate is given by
the singular value $\lambda_j = |\tilde{\lambda}_j|$
as suggested by the general discussion in the previous section.

Using \eqref{solution-z-y},
the integral on the deformed contour is written as 
\aln{
  Z&=\int_{\Sigma_\tau}d^Nz\,\exp\left(\frac{i}{2}
  \sum_{j=1}^N \tilde{\lambda}_j \, z_j^2\right)
\label{eq: harmonic-Z-flowed}\\
&=  \int_{\mathbb{R}^N}d^Ny \ \mathrm{det} \, J  \, 
\exp\left(- \frac{1}{2}
\sum_{j=1}^N \tilde{\lambda}_j \Bigl\{\sinh(2\tilde{\lambda}_j\tau)
- i\Bigr\} \, y_j^2\right) \ .
\label{eq: harmonic-Z-flowed-x}
}
Here
the Jacobi matrix $J$ associated with the deformation of the contour is given by
\aln{
  J_{jk}&=\frac{\partial z_j(y,\tau)}{\partial y_k}
  = \{ \cosh(\tilde{\lambda}_j\tau)+ i\sinh(\tilde{\lambda}_j\tau) \} \, \delta_{jk} \ ,
  \label{Jacobi-matrix-hos}
}
which is independent of $y$ and hence can be factored out of the integral.

From \eqref{eq: harmonic-Z-flowed-x},
one can see how the sign problem is solved by increasing $\tau$. 
The real part of the argument of the exponential function implies
that the region of $y_j$ that contribute to the integral is $|y_j| \lesssim \Delta_j $,
where
\begin{align}
  \Delta_j =
  \Bigl\{ \tilde{\lambda}_j \sinh(2\tilde{\lambda}_j\tau) \Bigr\}^{-1/2} \ .
  \label{def-Delta}
\end{align}
In that region, the phase factor of the integrand is close to unity as far as
\begin{align}
  \frac{1}{2} \, |\tilde{\lambda}_j| (\Delta_j)^2 \ll \pi \ .
    \label{sign-solved-each-mode}
\end{align} 
Combining \eqref{def-Delta} and \eqref{sign-solved-each-mode},
the sign problem associated with the eigenmode $j$ is solved if
\begin{align}
  \tau \gg \frac{1}{2 \lambda_j} \, \mathrm{arcsinh} \left( \frac{1}{2\pi} \right) \ .
  \label{sign-solved-each-mode-cond}
\end{align} 
Thus, in order to solve the sign problem of the whole system,
one has to satisfy \eqref{sign-solved-each-mode-cond}
for the smallest singular value $\lambda_{\rm min}$.
In that case, the magnification factor \eqref{magnification-factor}
for the largest singular value $\lambda_{\rm max}$ becomes
\begin{align}
  R_{\rm max}(\tau) & \sim \exp \left( c \, \frac{\lambda_{\rm max}}{\lambda_{\rm min}}\right) \ ,
  \label{magnification-factor-max}
  \\       c & \gg  \frac{1}{2} \, \mathrm{arcsinh}
  \left( \frac{1}{2\pi} \right) \sim 0.08 \ .
  \end{align}
When the singular value has a large hierarchy 
$\frac{\lambda_{\rm max}}{\lambda_{\rm min}} \gg 1 $,
$R_{\rm max}(\tau)$ can easily diverge,
which is nothing but the hierarchical growth problem.

\begin{figure}[t]
\begin{center}
 \includegraphics[width=3.4in]{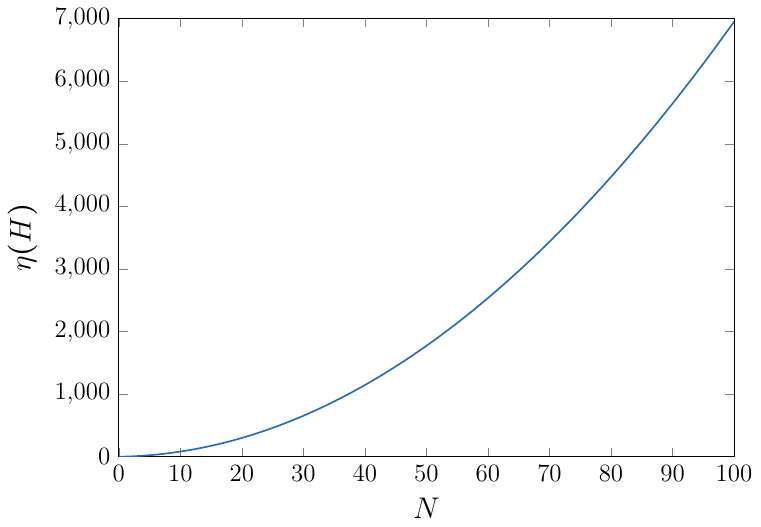}
 \caption{The condition number $\eta(H)$ given by \eqref{eq: eta-freeHessian}
   in the case of a harmonic oscillator is plotted against $N$
   for $m^2=1$ and $T=2$, which shows quadratic growth at large $N$.}
\label{fig: cond-num-harmonic}
\end{center}
\end{figure}

The ratio $\eta(H) \equiv \frac{\lambda_{\rm max}}{\lambda_{\rm min}}$
represents
the condition number of the matrix $H$.
Let us evaluate this quantity
explicitly in the case of a harmonic oscillator
using \eqref{lambda-hos}.
For $mT < \pi$, all the eigenvalues become positive $\tilde{\lambda}_j>0$
for sufficiently large $N$. In that case, $\eta(H)$ can be easily evaluated as
\aln{
\eta(H)&=\frac{4(N+1)^2\sin^2\frac{N\pi}{2(N+1)}-(mT)^2}
{4(N+1)^2\sin^2\frac{\pi}{2(N+1)}-(mT)^2}
\label{eq: eta-freeHessian}\\
& \sim
\frac{4}{\pi^2-(mT)^2} \, N^2 \ ,
}
in the $N\rightarrow \infty$ limit with fixed $m^2$ and $T$,
which corresponds to the continuum limit.
In Fig.~\ref{fig: cond-num-harmonic},
we plot $\eta(H)$ against $N$ for $m^2=1$ and $T=2$.
By plugging these values in \eqref{magnification-factor-max}, one finds that
$R_{\rm max}$ is as large as $2.4\times 10^{9}$
even for $N = 20$
used in our numerical simulation in
Section \ref{sec:reproduction-wave}.



%

Let us also note that the 
Jacobian that appears in \eqref{eq: harmonic-Z-flowed-x}
has a modulus that can be written
in terms of the magnification factor
$R_j(\sigma)$ defined in \eqref{magnification-factor} as
\aln{
  | \mathrm{det} \, J |
  &=\prod_{j=1}^N R_j(\tau)
  \sim e^{\Lambda \tau} \ ,
\label{eq: jacobi-growth}
}
where $\Lambda = \sum_{j=1}^N \lambda_j$ grows quadratically
with $N$ in the continuum limit.
In the present example of a harmonic oscillator,
the Jacobian is a constant that does not depend on the variables $y_j$,
which simply factors out of the integral in \eqref{eq: harmonic-Z-flowed-x}
as we already mentioned.
This is not the case
in a general system, however.
The Jacobian with
a huge modulus depending on the configuration
can cause an overlap problem in the on-axis approach as we discussed
in the Introduction.


In the example discussed above,
the origin of the large
$\eta(H)$ is the ratio of the momenta in the UV and IR
regions as one can see from \eqref{eq: eta-freeHessian}.
This is closely related to the recent discussion
that the flow equation in the continuum theory
does not allow a well-behaved solution due to the high frequency
modes \cite{Feldbrugge:2022idb}. 
In order to solve this problem,
it was proposed to
modify the real part of the action that appears
on the right-hand side of the flow equation
in the vicinity of the saddle points.

As we have seen in Section \ref{sec:hierarchical-prop},
a similar problem occurs
generally in a discretized theory with
a large number
$N$ of variables,
which
typically exhibits
a large hierarchy in the singular value spectrum of the Hessian.
We solve this problem
by
modifying the flow equation so that the growth rate of each mode becomes
identical,
while keeping the crucial property of the flow intact.





\section{A solution to the hierarchical growth problem}
\label{sec: precond}

Our basic strategy to solve the hierarchical growth problem
is to normalize the growth rates at each step of the flow.
For instance, in the example of a harmonic oscillator discussed in the
previous section,
we can replace \eqref{eq: flow-of-z} by
\aln{
  \frac{\partial z_j(y,\sigma)}{\partial \sigma}
  &=i \, \frac{\tilde{\lambda}_j}{|\tilde{\lambda}_j|} \,
  \overline{z_j(y,\sigma)} \ ,
\label{eq: flow-of-z-modified}
}
without spoiling the crucial property \eqref{eq: flow-essence}
of the flow equation.
Thus the hierarchical growth problem can be completely solved.
What we aim to do here is to generalize this prescription to an
arbitrary system.

\subsection{Preconditioned gradient flow equation}
  
Let us first note that the crucial property \eqref{eq: flow-essence}
of the gradient flow equation holds for
more general flows described by\footnote{Recently, it has been found that
  introducing a kernel in the complex Langevin equation is
  useful in stabilizing simulations in the CLM for real-time quantum
  evolution \cite{Alvestad:2022abf,Boguslavski:2022dee,Alvestad:2023jgl}.}
\aln{
\frac{d z_j}{d \sigma}
&= \mathcal{A}_{jk}(z,\bar{z}) \, \overline{\frac{\partial S}{\partial z_k}} \ ,
\label{eq: precond-flow}
}
with $\mathcal{A}(z,\bar{z})$ being an arbitrary positive-definite hermitian matrix,
which is not necessarily a holomorphic function of $z$.
This can be proved easily as
\aln{
\frac{d S(z(\sigma))}{d \sigma}
=\frac{\partial S}{\partial z_j}\mathcal{A}_{jk}(z,\bar{z})
\overline{\frac{\partial S}{\partial z_k}}>0 \ .
}
Note that $\mathcal{A}$ changes not only the deformed contour $\Sigma_\tau$
but also the thimbles $\Sigma_\infty$.
However, since the crucial property of the flow is kept intact,
$\text{Im}\,S$ is constant on each thimble, which implies that
the sign problem can be solved in the same way.
We discuss this point
in Appendix \ref{sec: thimbles} with an explicit example.

For the generalized flow \eqref{eq: precond-flow},
we find that the displacement
$\zeta(\sigma) \equiv z(x+\delta x,\sigma)-z(x,\sigma)$
of the two points after some flow time $\sigma$
is given by the equation
\begin{equation}
  \frac{d}{d \sigma} \zeta_i(\sigma)
  =\mathcal{A}_{ik} \overline{H_{kl}(z(\sigma))\, \zeta_l(\sigma)}
  + \left(
  \frac{\partial \mathcal{A}_{il}}{\partial z_k} \zeta_k(\sigma)
  + \frac{\partial \mathcal{A}_{il}}{\partial \bar{z}_k} \overline{\zeta_k(\sigma)} \right)
    \overline{ \frac{\partial S(z(\sigma))}{\partial z_l}} 
 \ ,
      \label{eq:zeta_floweq-preconditioned}
\end{equation}
instead of
the original one \eqref{eq: flow-of-displacement}.
Let us here assume that the first term is
dominant\footnote{This assumption is valid when $z(\sigma)$ is
  close to a saddle point, for instance. Otherwise, it should be 
simply regarded as a working hypothesis.}
in \eqref{eq:zeta_floweq-preconditioned}.
Then
the solution to \eqref{eq:zeta_floweq-preconditioned}
can be written formally as \eqref{zeta-formal-sol}
with ${\cal H}(\sigma)$ replaced by\footnote{Note that
  $\widetilde{\cal H}(\sigma)$ is not Hermitian in general
  unlike ${\cal H}(\sigma)$.}
\begin{align}
\widetilde{\cal H}(\sigma) &= 
  \begin{pmatrix}
    &
    {\cal A} \, \overline{H(z(\sigma))}  \\
        \bar{\cal A} \, H(z(\sigma))  &   
  \end{pmatrix} \ .
\label{def-cal-H-precond}
\end{align}
Recalling
the singular decomposition \eqref{eq:H-SVD} of $H(z(\sigma))$,
we can choose
\begin{align}
  \mathcal{A} = U(\sigma)^\dag \, \Lambda^{-1}(\sigma)\, U(\sigma) \ ,
  \label{eq:A-optimal}
\end{align}
so that
the problematic hierarchy of singular values 
in $H(z(\sigma))$ is completely eliminated as
\begin{align}
\bar{\cal A} \, H(z(\sigma))
&= U^\top (\sigma) \, U(\sigma) \ .
\end{align}
In this case, the $2N \times 2N$ matrix $\widetilde{\cal H}(\sigma)$
is again Hermitian and it has $N$ eigenvalues of $1$ and $-1$, respectively;
namely the growth rate of each eigenmode becomes identical.
(See \eqref{diagonalize-cal-H}.)
From this point of view, \eqref{eq:A-optimal} can be regarded
as the optimal choice for the ``preconditioner'' $\mathcal{A}$ 
in the generalized
flow equation \eqref{eq: precond-flow}.

In order to implement this idea in practice,
let us note 
that \eqref{eq:A-optimal} can be written
as\footnote{
  When $H$ has zero singular values, the preconditioner $\mathcal{A}$
  has to be regularized, for instance,
  as $\mathcal{A}=(\bar{H}(\bar{z})H(z)+\varepsilon)^{-1/2}$
  with a small positive $\varepsilon$.
  This is needed
  when one treats a system with symmetries. 
  While the deformed contour changes with $\varepsilon$, the integral
  remains unaltered.
  \label{footnote:zero-SV}
} 
\begin{align}
  \mathcal{A}(z(\sigma),\overline{z(\sigma)}) 
  &=
  \Big\{ H^\dag(z(\sigma)) H(z(\sigma)) \Big\}^{-1/2}
  = \Big\{ \overline{H(z(\sigma))} H(z(\sigma)) \Big\}^{-1/2} \ .
  \label{eq:A-optimal-2}
  \end{align}
Here we use the rational
approximation
\begin{align}
x^{-1/2} 
&\approx
a_0  + \sum_{q=1}^Q \frac{a_q}{ x + b_q } \ ,
\label{eq:rational_approximation}
\end{align}
which can be made accurate for a wide range of $x>0$
with the real positive parameters $a_q$ and $b_q$
generated by the Remez algorithm.
%
Thus we obtain
\begin{align}
    \mathcal{A}(z,\bar z)
    &\approx a_0 \, \textbf{1}_N
    + \sum_{q=1}^Q a_q \, \Big\{ \overline{H(z)} H(z)+b_q\, \textbf{1}_N \Big\}^{-1} \ .
    \label{eq:rational_approximation-2}
\end{align}
The matrix inverse $(\bar HH+b_q\textbf{1}_N)^{-1}$ does not have to
be calculated explicitly since it only appears in the algorithm 
as a matrix that acts on some
vector, which allows us to
use an iterative method for solving a linear equation
such as the conjugate gradient (CG) method.
The factor of $Q$ in the computational cost
can be avoided by the use of a multi-mass CG
solver \cite{Jegerlehner:1996pm,Jegerlehner:1997rn}.
These techniques are well known 
in the so-called Rational HMC
  algorithm \cite{Kennedy:1998cu,Clark:2006wq},
which is widely used in
  QCD with dynamical strange quarks \cite{Clark:2004cp} and
  supersymmetric theories such as the BFSS and IKKT matrix 
models (See Refs.~\cite{Catterall:2007fp,Anagnostopoulos:2007fw,Kim:2011cr}, 
for example.).

\subsection{Calculation of the Jacobian}
\label{sec:calc-jacobian}

In the GTM, one has to calculate
the Jacobian associated with the change of variables
defined by the gradient flow.
In the on-thimble approach, only the phase factor of the Jacobian
has to be reweighted,
while in the on-axis approach,
not only the phase factor but also the modulus has to be reweighted.
The calculation of the Jacobian
gets modified when one introduces the preconditioner
to the flow equation.
In this subsection, we discuss how this can be done efficiently.

Let us note first that
the flow of the Jacobi matrix is given
by\footnote{The flow equation \eqref{eq:zeta_floweq-preconditioned}
  for the displacement
  can be obtained
from \eqref{eq:Jacobian_floweq-preconditioned} by using
$\zeta_j(\sigma) = J_{jk}(\sigma) \, \delta x_k$.
} 
\begin{equation}
    \frac{d}{d \sigma}J_{ij}(\sigma)
  =\mathcal{A}_{ik} \overline{H_{kl}(z(\sigma))\, J_{lj}(\sigma)}
  + \left(
  \frac{\partial \mathcal{A}_{il}}{\partial z_k} J_{kj}(\sigma)
  + \frac{\partial \mathcal{A}_{il}}{\partial \bar{z}_k} \overline{J_{kj}(\sigma)} \right)
    \overline{ \frac{\partial S(z(\sigma))}{\partial z_l}} 
 \ ,
      \label{eq:Jacobian_floweq-preconditioned}
\end{equation}
for the preconditioned flow equation \eqref{eq: precond-flow}.
(The corresponding flow for the original flow equation \eqref{eq: standard-flow}
can be retrieved by setting $\mathcal{A}$ to an identity matrix.)
Using the expression \eqref{eq:rational_approximation-2},
the derivative of $\mathcal{A}$
in \eqref{eq:Jacobian_floweq-preconditioned}
can be calculated as
\begin{align}
  \frac{\partial\mathcal{A}}{\partial z_k}
  &=-\sum_{q=1}^Qa_q (\bar HH+b_q\textbf{1}_N)^{-1}\bar H 
  \frac{\partial H}{\partial z_k}
  (\bar HH+b_q\textbf{1}_N)^{-1} \ ,
  \nn
  \\
  \frac{\partial \mathcal{A}}{\partial \bar{z}_k}
  &=-\sum_{q=1}^Qa_q(\bar HH+b_q\textbf{1}_N)^{-1}
  \overline{ \frac{\partial H}{\partial z_k} }
  H(\bar HH+b_q\textbf{1}_N)^{-1}  \ .
  \label{del-A-expressions}
\end{align}
Note that
the matrix inverse $(\bar HH+b_q\textbf{1}_N)^{-1}$ appears twice
in the above expressions.
In order to use
the idea of the multi-mass solver,
which enables us to avoid repeating the CG procedure $Q$ times
(See the explanation below
Eq.~\eqref{eq:rational_approximation-2}.),
we have to apply the matrix inverse to some different vectors separately.
For that, we regard the matrix equation
\eqref{eq:Jacobian_floweq-preconditioned}
as a set of vector equations for $i=1 , \cdots , N$.
We can calculate the right-hand side of the vector equation for
a particular $i=I$ by multiplying a unit vector $e_i^{(I)}=\delta_{iI}$
to the right-hand side of \eqref{eq:Jacobian_floweq-preconditioned}.
Thus
the matrix inverse
$(\bar{H}H+b_q{\bf 1})^{-1}$ that appears
in
\eqref{del-A-expressions}
can be applied to
the two vectors $e_i^{(I)}$ and $\overline{\partial _l S}$
separately.

In fact, the expressions \eqref{del-A-expressions} have to be used
also when we compute the force term in the HMC algorithm
in the on-axis approach. (See Appendix \ref{sec: HMC-on-axis-review}
for a brief review.)
In that case, too, we can apply the idea of the multi-mass solver
thanks to the structure of calculations in the backpropagation
as we discuss in Appendix \ref{sec: forces}.

The calculation of the Jacobi matrix that we discussed above
requires O($N^2$) computational cost for a local action
since it involves multiplications of a sparse matrix 
to a dense matrix\footnote{For a non-local action,
  the cost increases
  by a factor of O($N$) since the sparseness is lost.\label{footnote:non-local}}.
On the other hand,
the calculation of its determinant
requires O($N^3$) computational cost,
which is not affected by the preconditioning at all.
These calculations appear only in the reweighting procedure,
which can be done off-line after generating configurations,
and the procedure can be
  parallelized trivially without any overhead due to
  communications.
On the other hand,
all the calculations
for generating configurations
require
only O($N$) computational cost for a local action
even after implementing the preconditioner in the gradient flow equation
since it just involves multiplications of a sparse matrix to a
vector.

In passing,
let us also mention that
there is actually
a cheaper preconditioner,
which can be obtained by considering only the free part of the action
when we calculate the Hessian $H$ to be used in \eqref{eq:A-optimal-2}.
Since the preconditioner $\mathcal{A}(z,\bar{z})$ in this case becomes
a constant matrix independent of the configuration,
we only have to compute $\mathcal{A} $
and diagonalize it once and for all.
The rational approximation \eqref{eq:rational_approximation-2}
is not needed any more.
This cheaper preconditioner works efficiently in the weak coupling regime,
and the computational cost can be reduced typically by
an order of magnitude compared to the full preconditioner.


\section{Demonstration of the preconditioned flow}
\label{sec: simulation}

In this section, we demonstrate
how the preconditioned flow equation works
in the application of GTM to the real-time quantum evolution
of an anharmonic oscillator.
Here we use the on-axis approach for the HMC algorithm,
where the force term is calculated by
the backpropagation \cite{Fujisawa:2021hxh}.
In particular,
we show that
the long autocorrelation in the generated
configurations
is drastically reduced by preconditioning the flow.
We also show that the preconditioning
drastically reduces the modulus of the Jacobian
and its fluctuations,
thereby solving the overlap problem in the on-axis approach.
%
%

\subsection{Simulation setup}
\label{sec:sim-setup}

The system we deal with in this section
is the same as
\eqref{eq: harmonic-Z} except that
we consider an anharmonic potential
\begin{align}
V(x) &=
\frac{1}{4!} \, \lambda \, x^4 \ .
    \label{eq:def-V-anharmonic}
\end{align}
Here we introduce the initial wave function
\begin{align}
  \psi_{\rm i} (x) \propto\exp \left\{-\frac{1}{4}\gamma \,
  (x-x_{\rm i})^2\right\} \ ,
    \label{eq:initial-wf}
\end{align}
and calculate the time-evolved wave function $\psi_{\rm f}(x)$
after some time $T$,
which is given by the path integral \eqref{eq:general-path-integral}
up to some known normalization factor
using the action
\aln{
S(x;x_{\rm f},x_{\rm i})=
&-i\sum_{j=1}^N\epsilon\left\{\frac{1}{2} \left(
\frac{x_{j+1}-x_j}{\epsilon} \right)^2
-  \frac{V(x_{j+1}) + V(x_j)}{2}\right\}
+\frac{1}{4} \, \gamma \, (x_1-x_{\rm i})^2\ ,
\label{eq:action-anHOS}
}
where $T=N\epsilon$ and $x_{N+1} \equiv x_{\rm f}$.
Similar calculations have been
done
using the original flow
in Ref.~\cite{Fujisawa:2021hxh}
although $N$ had to be restricted to small values such as $N=9$.
The main point here is that the preconditioned flow enables us
to increase $N$ without
any problems.


In order to avoid the ergodicity problem
concerning
multiple thimbles, we integrate over the flow time \cite{Fukuma:2020fez}
as described in Ref.~\cite{Fujisawa:2021hxh}
in the on-axis approach.
The Hamiltonian to be used in the fictitious Hamilton dynamics
of the HMC algorithm is
given by
\aln{
  H=\frac{1}{2m^2(\tau)}\sum_{j=1}^Np_j^2
   + \frac{1}{2\tilde{m}^2} \, p_\tau^2
  + {\rm Re} \,S(z(x,\tau)) +  W(\tau) \ ,
\label{eq: Hamiltonian}
}
where $p_j$ and $p_\tau$ are the conjugate momenta corresponding
to $x_j$ and the flow time $\tau$, respectively.
The function $m(\tau)$ 
and the parameter $\tilde{m}$
are introduced for optimization as we explain shortly.
The total time $s_{\rm f}$ and the step size $\Delta s$
for the fictitious Hamilton dynamics are the parameters of the HMC algorithm
that can be optimized in a standard manner.
In all our simulations,
we choose $s_{\rm f}=1$ and $\Delta s=0.05$,
and the number of time steps for solving the flow equation is set to
$N_\tau=10$.

The parameter $\tilde{m}>1$ in \eqref{eq: Hamiltonian} is introduced
only for simulations with the original flow
in order to avoid the force in the $\tau$-direction getting too large depending on $x$,
which causes the drop of the acceptance rate.
We have increased $\tilde{m}$ from unity until the acceptance rate becomes
reasonably high.


On the other hand, the function $m(\tau)$ 
in \eqref{eq: Hamiltonian}
is chosen to be
the typical largest singular value of the Jacobi matrix $J(x,\tau)$
based on
the discussion in Section \ref{sec: subtlety}
so that the eigenmode
with the largest growth rate does not diverge.\footnote{
  In Ref.~\cite{Fujisawa:2021hxh},
  $m(\tau)$ was chosen to be
  the typical value of $|\mathrm{det}\,J(x,\tau)|^\frac{1}{N}$,
which
corresponds to taking the geometric average of the growth rate.
However, the effective stepsize in this case 
becomes too large for the eigenmode with the largest growth rate,
which causes divergence during simulations at large $N$. 
} 
Then the hierarchical growth problem manifests itself as
long autocorrelation due to the modes with small growth rate.

The function $W(\tau)$
in \eqref{eq: Hamiltonian} is determined so that the distribution
of $\tau$ becomes flat within the region $\tau \in [\tau_{\rm min},\tau_{\rm max}]$,
where $\tau_{\rm max}$ has to be chosen to be
large enough to solve the sign problem
and $\tau_{\rm min}$ has to be chosen to be
small enough to solve the ergodicity problem.
The actual form of the function $W(\tau)$ as well as that of $m(\tau)$
is determined iteratively by improving them step by step
using the results obtained by the previous simulation.

%



\begin{figure}[t]
\begin{tabular}{cc}
\begin{minipage}[t]{.5\textwidth}
\includegraphics[width=3in]{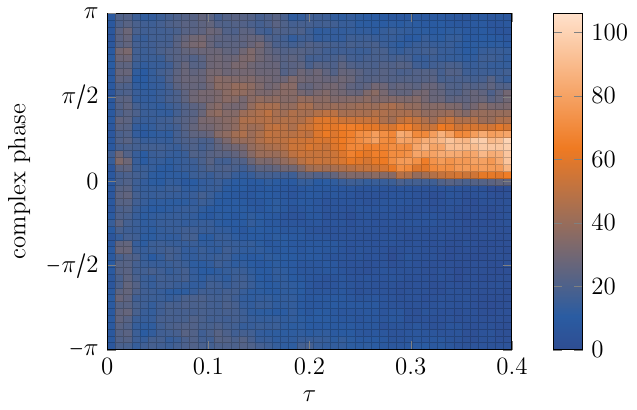}
\end{minipage}
\begin{minipage}[t]{.5\textwidth}
\includegraphics[width=3in]{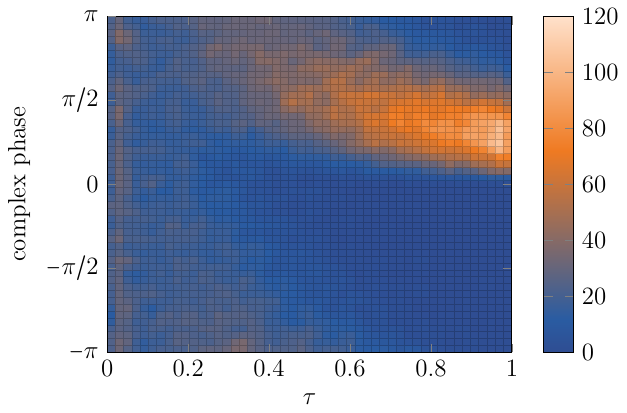}
\end{minipage}
\end{tabular}
\caption{The distribution of the phase of the integrand
  of \eqref{part-fn-deformed}
  and the flow time $\tau$ obtained by simulations
  is shown
  for the original flow (Left) and the preconditioned flow (Right)
  in the case of
  $N=6$, $T=2$, $\lambda=1$, $x_{\rm i}=1$, $\gamma=1$ and $x_{\rm f}=0$.
  The color indicates the number of
  configurations
  in each bin.
  The total number of configurations is 50000,
  and we have 50 bins in both the horizontal
  and vertical directions.
}
\label{fig: phases}
\end{figure}

Let us discuss how we
determine
the optimal region $[\tau_\text{min},\,\tau_\text{max}]$
of $\tau$ to be used in the simulation,
which actually depends on whether we use the preconditioned flow or not.
In Fig.~\ref{fig: phases},
we show the distribution
of the phase of the integrand in \eqref{part-fn-deformed}
including the phase coming from the Jacobian
for the original flow (Left) and the preconditioned flow (Right)
in the case of
$N=6$, $T=2$, $\lambda=1$, $x_{\rm i}=1$, $\gamma=1$ and $x_{\rm f}=0$.

The sign problem is solved in
the region of $\tau$ that exhibits some non-uniformity
in the distribution of the phase.
However, in order to avoid the ergodicity problem,
one should also sample the region of $\tau$
in which the distribution of the phase is almost uniform.
Based on these criteria,
we choose the range of $\tau$ to be
$[0.02,\,0.2]$ for the
original flow,
and 
$[0.4,\,0.8]$ for
the preconditioned flow in this case.

See Appendix \ref{sec: parameters} for
the choice of the parameters
$\tilde{m}$, $\tau_{\rm min}$, $\tau_{\rm max}$, $m(\tau)$ and $W(\tau)$
determined in the way described above for each case.

\subsection{Reduction of the autocorrelation} 

\begin{figure}[t]
\begin{center}
\includegraphics[width=2.9in]{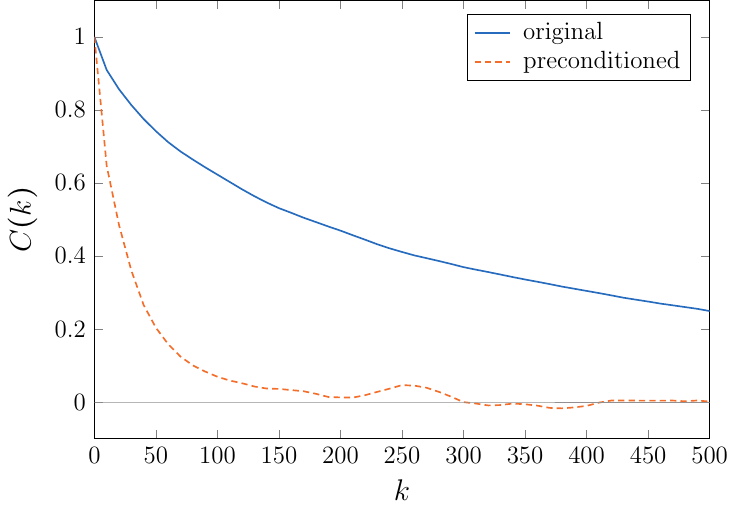}
\caption{The autocorrelation
  of ${\bf x}$
  is plotted against the number of separation
  using 50000 configurations
  obtained by simulations
  with $N=6$, $T=2$, $\lambda=1$, $x_{\rm i}=1$, $\gamma=1$ and $x_{\rm f}=0$.
  The blue solid line and the orange dashed line
  represent the results obtained by
  the original flow and the preconditioned flow, respectively.
}
\label{fig: auto}
\end{center}
\end{figure}

Let us first show that the autocorrelation is drastically
reduced by preconditioning the flow equation.
We define the autocorrelation
in the generated configurations ${\bf x} = (x_1 , \cdots , x_N )$
as follows.
Let ${\bf x}^{(k)}$ ($k = 1, \cdots , n$)
be the $k$-th configuration.
We denote the average of the $n$ configurations as
$\overline{{\bf x}} = \frac{1}{n} \sum_{k=1}^n {\bf x}^{(k)}$.
Then the autocorrelation
can be defined as
\aln{
  C(k)
  \equiv \frac{1}{v} \frac{1}{n-k}\sum_{m=1}^{n-k}
( {\bf x}^{(m)}-  \overline{\bf x} )
\cdot ({\bf x}^{(m+k)}- \overline{\bf x} ) \ ,
\label{eq: auto}
}
where $v$ is the variance defined by
\aln{
  v = \overline{  {\bf x} \cdot {\bf x} }
  -   \overline {{\bf x}} \cdot  \overline {{\bf x}}    \ .
\label{eq: def-variance}
}


In Fig.~\ref{fig: auto},
we show the autocorrelation
of ${\bf x}=(x_j)$ obtained by simulations
with
the same parameters as in Fig.~\ref{fig: phases}.
The region of $\tau$ is chosen to be
$[0.02,\,0.2]$ and $[0.4,\,0.8]$
for the original flow and the preconditioned flow, respectively.
We find that the autocorrelation is reduced by a factor of 10
for the preconditioned flow, which suggests
that the hierarchical growth problem is avoided.


\subsection{Reduction of the modulus of the Jacobian}

\begin{figure}[t]
\begin{tabular}{lcr}
\begin{minipage}[t]{.5\textwidth}
\centering\includegraphics[width=2.7in]{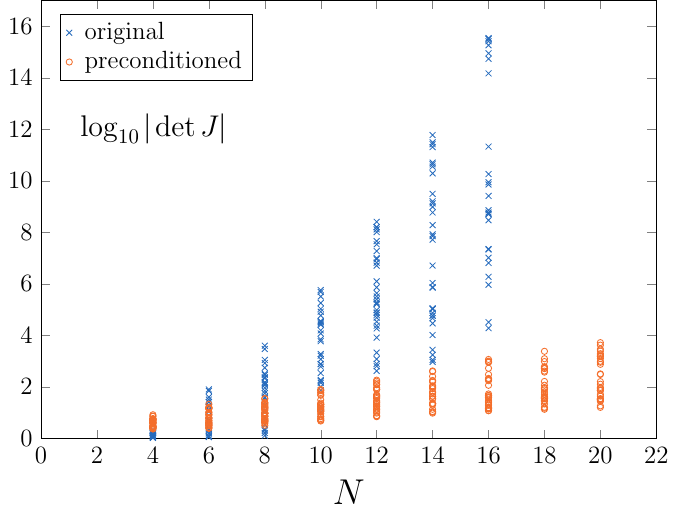}
\end{minipage}
\begin{minipage}[t]{.5\textwidth}
  \centering\includegraphics[width=2.7in]{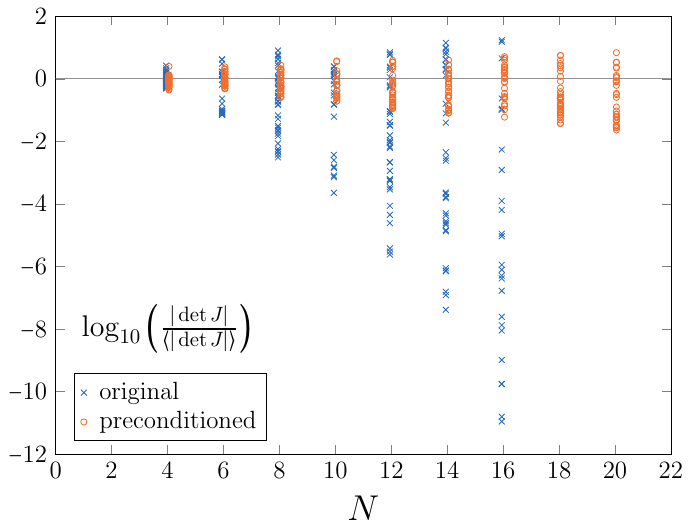}
\end{minipage}
\end{tabular}
\caption{(Left) The scattered plots
  of the modulus $|\mathrm{det} J|$ of the Jacobian obtained by simulations
is shown for various $N$ ($4 \le N \le 20$)
in the case of $T=2$, $\lambda=1$, $x_{\rm i}=1$, $\gamma=1$ and $x_{\rm f}=0$.
The blue crosses and the orange circles represent the results
for the original flow and the preconditioned flow, respectively.
For each $N$, we have used 30 configurations obtained within 10000 trajectories.
(Right) Similar plots for the normalized modulus
$|\mathrm{det} J|/\langle|\mathrm{det} J|\rangle$
of the Jacobian
are shown for various $N$.
}
\label{fig: Jacobian-modulus}
\end{figure}

Next we discuss the Jacobian associated with the change of variables
by the gradient flow, which can typically have a large modulus
as we mentioned below \eqref{eq: jacobi-growth} in Section \ref{sec:hos-example}.
This, in particular, causes the overlap problem in the on-axis approach.
We will see below that the preconditioned flow reduces
the modulus of the Jacobian and its fluctuation drastically,
thereby solving this problem as well as the hierarchical growth problem.

In
Fig.~\ref{fig: Jacobian-modulus} (Left),
we show the scattered plot of $\log_{10}|\mathrm{det} J|$
obtained by simulations
for various $N$ ($4 \le N \le 20$)
with $T=2$, $\lambda=1$, $x_{\rm i}=1$, $\gamma=1$ and $x_{\rm f}=0$.
We have fixed the range of $\tau$ as
$[0.02,\,0.2]$ and $[0.4,\,0.8]$ 
for the cases with and without preconditioning, respectively. 
The blue crosses and the orange circles
represent the results for the original flow
and the preconditioned flow, respectively.
%
%
%
%
Our results show that 
the preconditioning reduces the values of
$|\mathrm{det} J|$ drastically as expected.
%
Furthermore, from Fig.~\ref{fig: Jacobian-modulus} (Right),
we find that the preconditioning suppresses
the fluctuation of $|\mathrm{det} J|$ drastically,
which suggests that it can
also solve the overlap problem in the on-axis approach.

Without preconditioning,
it was not even possible to
perform simulations for $N>16$
since the typical magnitude of the HMC force in the $\tau$-direction
changes too much with $x$, and we were not able to control
it by the $x$-independent potential
$W(\tau)$.
This may be viewed as
another manifestation of the hierarchical growth problem.
Such behaviors were
not seen in simulations with
the preconditioned flow.

\subsection{Result for the time-evolved wave function}
\label{sec:reproduction-wave}

\begin{figure}[t]
\begin{center}
\includegraphics[width=3.4in]{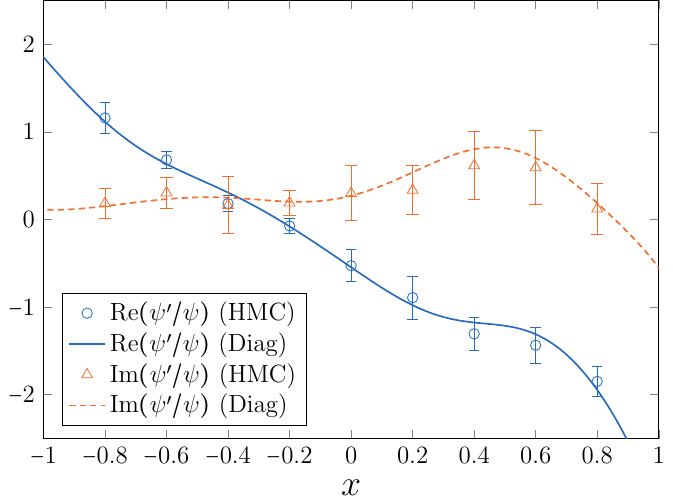}
\caption{The quantity \eqref{eq: observable} derived from
  the time-evolved wave function is shown for
  $N=20$, $T=2$, $\lambda=30$, $x_{\rm i}=0.3$ and $\gamma=4$.
  The blue circles and the orange triangles
  represent the real part and the imaginary part, respectively.
  Each point is obtained by taking an average over
  30000 configurations generated by an independent simulation using $x_{\rm f}=x$.
  The blue solid line (real part) and the orange dashed line (imaginary part)
  represent
  the results
  obtained by solving the Schr\"odinger equation with Hamiltonian diagonalization.
}
\label{fig: final-WF}
\end{center}
\end{figure}

Finally we show that the preconditioned flow enables us
to obtain
results for
the time-evolved wave function $\psi_{\rm f}(x_{\rm f})$
even at large $N$ and at strong coupling (large $\lambda$).
As has been done in Ref.~\cite{Fujisawa:2021hxh},
we calculate
\aln{
  \frac{\partial}{\partial x_{\rm f}}&\log(\psi_f(x_{\rm f}))
  =- \left \langle \frac{\partial S}{\partial x_{\rm f}} \right \rangle
  =\left\langle  i\left(\frac{x_{\rm f}-x_N}{\epsilon}
   - \frac{\epsilon }{2} \, V'(x_{\rm f} )\right)\right\rangle \ ,
\label{eq: observable}
}
which is directly accessible by calculating
the expectation value on the right-hand side.
In Fig.~\ref{fig: final-WF},
we show our results for
$N=20$, $T=2$, $\lambda=30$, $x_{\rm i}=0.3$ and $\gamma=4$.
For each $x_{\rm f}$, we have determined the range of $\tau$
as presented in Appendix \ref{sec: parameters}.
Our results are in good agreement with the results obtained by
solving the Schr\"odinger equation with Hamiltonian diagonalization.

%
%
%

In Ref.~\cite{Fujisawa:2021hxh}, it was difficult to obtain reliable results
as one goes
beyond $N=9$
because of the overlap problem
even at $\lambda=1$.
Here we are able to obtain results for $N=20$
%
even at strong coupling $\lambda=30$ 
without any problems.



\section{Summary and discussions}
\label{sec: summary}

In this paper we have pointed out a problem in the gradient
flow equation, which is used
in the GTM to deform the integration contour into
the complex plane. The property of the flow
that plays an important role in solving the sign problem
is that it maps a small region on the
real axis to a region on the deformed contour
which becomes exponentially large with the flow time.
The problem is that the growth rate for each mode
typically has a huge hierarchy
when the system size becomes large.
If one chooses the flow time to be large enough to solve the sign problem
associated with the slowly growing modes,
the fast growing modes tend to diverge.

In order to solve this hierarchical growth problem,
we have proposed to modify the flow equation
by the preconditioner, which makes the growth rate equal
without spoiling the crucial properties of the flow equation.
This preconditioner can be implemented practically in the GTM
with the standard techniques used in the Rational HMC algorithm.

We applied this method to the 
real-time quantum evolution
of an anharmonic oscillator
using the on-axis approach with backpropagation for calculating the
HMC force \cite{Fujisawa:2021hxh}.
In the on-axis approach, the hierarchical growth problem manifests itself in the long
auto-correlation time since one cannot choose the parameters
in the HMC algorithm for each mode separately.
Our results indeed show that the auto-correlation time is reduced drastically.
Moreover, the modulus of the Jacobian, which has to be taken into account
by reweighting, is also reduced drastically.
This solves the overlap problem in reweighting, which is caused
by large fluctuations of the modulus of the Jacobian.
Thus we were able to simulate the system with the size $N$
that was not accessible
without preconditioning the flow \cite{Fujisawa:2021hxh}.
Note also that
this is achieved with a very strong coupling $\lambda=30$ in the
quartic potential \eqref{eq:def-V-anharmonic}, which makes the
flow equation highly non-linear.

As a side remark,
the preconditioned flow
solves
yet another problem of
the original gradient flow, which occurs
when
the integrand has zeroes \cite{Tanizaki:2017yow}.
In that case, there are points on the real axis that flow
into the zeroes within finite flow time,
and the right-hand side of the flow equation
blows up
as one approaches the zeroes.
A practical solution proposed in Ref.~\cite{Tanizaki:2017yow}
is to
add some regulator to the flow equation,
which modifies the flow only in the vicinity of the zeroes.
We consider that
this problem does not occur for the preconditioned flow
since the preconditioner cancels the divergence of the right-hand side
that occurs as one approaches the zeroes.
See Appendix \ref{sec: blow-up} for 
discussions on this issue.
%
%

While we focused on the on-axis approach in this paper,
the hierarchical growth problem of the flow
may also 
affect the on-thimble approach.
There, we have to solve a set of
equations iteratively by Newton's method
in order to make sure that the updated configurations
are constrained
on the deformed contour. 
This procedure appears at every step of the fictitious time evolution
in the HMC algorithm.
The coefficient matrix of these equations
involves the Jacobian, which has a huge condition number for the original flow.
Therefore, the solution of the equations may
suffer from large numerical errors and/or slow convergence
when the system size becomes large.
The preconditioned flow may be useful in solving such problems.
(See Ref.~\cite{Fukuma:2023eru} for a new proposal related to this issue.)


As we have shown in this paper,
the preconditioned flow enables us to apply
the GTM to much larger systems than those accessible with the original flow.
In particular, it gives us an access to the continuum limit
and to the strong coupling regime,
which enabled us to establish a new picture of quantum tunneling
in the real-time path integral formalism \cite{Nishimura:2023dky}.
We expect that there are many other applications awaiting us to explore.

\subsection*{Acknowledgements}
We would like to thank Yuhma Asano, Genki Fujisawa,
Masafumi Fukuma and Nobuyuki Matsumoto
for valuable discussions.
The computations were carried out on
the PC clusters in KEK Computing Research Center
and KEK Theory Center.
K.S.\ was supported by the Grant-in-Aid for JSPS Research Fellow,
No.\ 20J00079.
A.Y.\ is supported by a Grant-in-Aid for Transformative Research Areas
``The Natural Laws of Extreme Universe---A New Paradigm for Spacetime and
Matter from Quantum Information'' (KAKENHI Grant No.\ JP21H05191) from
JSPS of Japan.

\appendix

\section{The absence of the blow-up problem}
\label{sec: blow-up}

Here we discuss
a problem that occurs when the integrand of the partition function has zeroes.
In such a case, the gradient flow can reach the zeroes within finite flow time
and the right-hand side of the flow equation blows up.
This is
the blow-up problem of the GTM, which was discussed in Ref.~\cite{Tanizaki:2017yow}.
Let us first emphasize that
this problem occurs
although the partition function is totally well defined.
The deformed contour one obtains
at sufficiently long flow time
is nothing but a set of contours connected at the
zeroes\footnote{See Ref.~\cite{Feldbrugge:2023mhn}
for recent discussions on this case.} .
It is therefore just a technical problem of how
to deal with the divergence that occurs while solving the flow equation.
Here we argue that this problem
is naturally avoided in the preconditioned flow
since
it
slows down the flow as one approaches the zeroes.


To simplify the argument, we consider a single-variable case $N=1$
in \eqref{eq:general-path-integral} with real $S(x)$.
In this case, the original flow equation \eqref{eq: standard-flow} reads
\aln{
  \frac{d x(\sigma)}{d \sigma}
  &= S'(x(\sigma))  \ ,
\label{eq: standard-flow-one-var}
}
whereas the preconditioned flow \eqref{eq: precond-flow} with
\eqref{eq:A-optimal-2} reads
\aln{
  \frac{d x(\sigma)}{d \sigma}
  &=
\frac{S'(x(\sigma))}{|S''(x(\sigma))|} \ ,
\label{eq: precond-flow-one-var}
}
where ${}'$ represents the derivative with respect to $x$.

For instance, let us consider the case with
\begin{align}
  S(x) &= \frac{1}{2(n+1)} \, x^{2(n+1)} \ , \qquad n=1,2,\cdots \ ,
    \label{eq:example-quartic}
\end{align}
which gives $e^{-S(x)}\rightarrow 0$ for $x \rightarrow \pm \infty$.
The saddle point is $x=0$, and there are flows towards the
singularities at $x = \pm \infty$.
Solving the original flow equation \eqref{eq: standard-flow-one-var}, one obtains
\begin{align}
  x(\sigma) &=
\left\{
\begin{array}{cc}
  \left\{ x(0)^{-2n} - 2n \sigma \right\}^{-\frac{1}{2n}} &
  \mbox{~for~}x(0) > 0 \ , \\
  - \left\{ x(0)^{-2n} - 2n \sigma \right\}^{-\frac{1}{2n}} &
  \mbox{~for~}x(0) < 0 \ , 
\end{array}
  \right.
\end{align} 
which reaches the singularities within finite flow time
$\sigma= \frac{1}{2n}x(0)^{-2n}$.
However, for the preconditioned flow \eqref{eq: precond-flow-one-var},
one obtains
\begin{align}
  x(\sigma) &= x(0) \, e^{\sigma/(2n+1)} \ , 
\end{align} 
which reaches the singularities only in the $\sigma \rightarrow \infty$ limit.


Next we consider the case with\footnote{Strictly speaking, the partition function
  is not finite
  in this case.
  In order to make it finite, one can add a term like
$\frac{1}{2} \epsilon x^2$ in the action, which clearly does not affect the
flow near the singularity $x=0$.}
\begin{align}
  S(x) &= - \log (x^{2n})  \ , \qquad n=1,2,\cdots \ ,
    \label{eq:example-log}
\end{align}
which gives $e^{-S(x)}\rightarrow 0$ for $x \rightarrow 0$.
The saddle point is $x=\pm \infty$, and there are flows towards the
singularity at $x = 0$.
Solving the original flow equation \eqref{eq: standard-flow-one-var}, one obtains
\begin{align}
  x(\sigma) &=
\left\{
\begin{array}{cc}
\sqrt{  x(0)^2 - 4n  \sigma }  & \mbox{~for~}x(0) > 0 \ , \\
-  \sqrt{  x(0)^2 - 4n  \sigma } & \mbox{~for~}x(0) < 0 \ , 
\end{array}
  \right.
\end{align} 
which reaches the singularity within finite flow time $\sigma= \frac{1}{4n}x(0)^{2}$.
However, for the preconditioned flow \eqref{eq: precond-flow-one-var},
one obtains
\begin{align}
  x(\sigma) &= x(0) \, e^{-\sigma} \ ,
\end{align} 
which reaches the singularity only in the $\sigma \rightarrow \infty$ limit.

Thus the preconditioning \eqref{eq: precond-flow-one-var}
makes the singularities unreachable within finite flow time.
While we have discussed a single variable case
with a real action
for simplicity,
this property of the preconditioned flow is considered to be quite general.
For instance, the quantum mechanical system with
the anharmonic potential \eqref{eq:def-V-anharmonic}
suffers from a blow-up problem for sufficiently long flow time
if one uses the original flow equation
similarly to the example \eqref{eq:example-quartic}.
We consider that the preconditioned flow naturally avoids this problem.
A more careful study on this issue shall be left for future investigations.

\section{Modification of the thimbles by preconditioning}
\label{sec: thimbles}

The preconditioning of the gradient flow equation changes the
deformed contour.
In fact, it also modifies the Lefschetz thimbles,
which appear as the deformed contour in the long flow time
$\tau \rightarrow \infty$ limit.
On the other hand, the preconditioning does not alter
the saddle points, which are defined by $\frac{\del S(z)}{\del z}=0$
independently of the flow.
In this section, we discuss these points
using a simple example. 


Let us consider
an integral \eqref{eq:general-path-integral}
with $N=2$ variables, where the action is given by 
\aln{
S(z)=\frac{1}{2}(z_1{}^2-iz_2{}^2) \ . 
}
The saddle point is given by $z_1 = z_2 = 0$.
The original flow equations read
\aln{
  \frac{d z_1}{d \sigma}
  =\bar{z}_1 \ , \quad \quad  \frac{\partial z_2}{\partial \sigma}=i\bar{z}_2 \ ,
  \label{original-flow-example}
}
which is linear.
Using
the real-variable notation 
$z_1=x_1+iy_1$, $z_2=x_2+iy_2$,
the flow equation \eqref{original-flow-example}
can be rewritten as
\aln{
  \frac{d}{d \sigma}
  \left(\begin{array}{c}x_1\\y_1\\x_2\\y_2\end{array}\right)
    =
    M
\left(\begin{array}{c}x_1\\y_1\\x_2\\y_2\end{array}\right) \ , \quad \quad
  M =
      \left(\begin{array}{cccc}
      \mbox{~}1 & ~ & \mbox{~~} & \mbox{~~} \\
      ~ & -1 & ~ & ~ \\
      ~ & ~ &  ~ & 1 \\
      ~ & ~ & 1 & ~ 
    \end{array}\right) \ .
}
The Lefschetz thimble $\Sigma_\infty$ is spanned by
the eigenvectors of the matrix $M$ corresponding to positive
(degenerate) eigenvalue $1$, which are given by
\aln{
v_1=\left(\begin{array}{c}1\\0\\0\\0\end{array}\right) \ , \quad \quad
v_2=\left(\begin{array}{c}0\\0\\1\\1\end{array}\right) \ .
}

Next we consider the generalized flow \eqref{eq: precond-flow}
with
a preconditioner\footnote{Note that
  the optimal choice is actually $\mathcal{A}= {\bf 1}$ in the present case.
  The aim here is to demonstrate that nontrivial $\mathcal{A}$ modifies
  the thimble.}
\aln{
  \mathcal{A}
  =\left(\begin{array}{cc}5&1\\1&5\end{array}\right) \ .
}
The preconditioned flow equations are given by
\aln{
  \frac{\partial z_1}{\partial \sigma}=5\bar{z}_1+i\bar{z}_2 \ ,
  \quad \quad
  \frac{\partial z_2}{\partial \sigma}=\bar{z}_1+5i\bar{z}_2 \ .
  \label{precond-flow-example}
}
Using the real-variable notation,
the preconditioned flow equation \eqref{precond-flow-example}
becomes
\aln{
  \frac{d}{d \sigma}
  \left(\begin{array}{c}x_1\\y_1\\x_2\\y_2\end{array}\right)
    =
    \widetilde{M}
\left(\begin{array}{c}x_1\\y_1\\x_2\\y_2\end{array}\right) \ , \quad \quad
  \widetilde{M} =
      \left(\begin{array}{cccc}
       \mbox{~}5 & ~ & \mbox{~~} & \mbox{~}1 \\
        ~ & -5 & 1 & ~ \\
      1 & ~ &  ~ & 5 \\
      ~ & -1 & 5 & ~ 
    \end{array}\right) \ .
}
The Lefschetz thimble $\widetilde{\Sigma}_\infty$ is spanned by
the eigenvectors of the matrix $\widetilde{M}$
corresponding to positive eigenvalues $4\sqrt{2}$ and $3\sqrt{2}$,
which are given, respectively, by
\aln{
\tilde{v}_1=\left(\begin{array}{c}5 + 4\sqrt{2}\\
5 - 3\sqrt{2}\\
1 + 5\sqrt{2}\\
7\end{array}\right) \ ,
\quad \quad
\tilde{v}_2=\left(\begin{array}{c}5 + 3\sqrt{2}\\
5 - 4\sqrt{2}\\
1 - 5\sqrt{2}\\
-7 \end{array}\right) \ .
}
Thus we find that the thimble
$\widetilde{\Sigma}_\infty$ is different from the original one $\Sigma_\infty$.
Note, however, that both thimbles are
embedded in a real 3-dimensional hypersurface
defined by
$\{(x_1~y_1~x_2~y_2)
~|~\text{Im}\,S=2x_1y_1-x_2^2+y_2^2=0\} \in\mathbb{R}^4$,
and therefore the sign problem is solved in both cases. 

\section{Application of the HMC using the on-axis approach}
\label{sec: HMC-on-axis-review}

In this section, we briefly review how one can apply
the HMC algorithm to the GTM using the on-axis approach.
The key idea that makes this feasible is the backpropagation
in calculating the force term in the fictitious Hamilton dynamics
of the HMC algorithm \cite{Fujisawa:2021hxh}.
Here we fix the flow time for simplicity, but
it is straightforward to implement
the integration over the flow time,
which is needed in avoiding the ergodicity problem.
(See Ref.~\cite{Fujisawa:2021hxh} for the details.)


Let us consider the general partition function given by
\eqref{eq:general-path-integral}.
The expectation value of the observable is defined by
\aln{
  \langle\mathcal{O}\rangle
  &=\frac{1}{Z}
  \int_{\mathbb{R}^N} dx\,e^{-S(x)}\mathcal{O}(x) \ .
  \label{observable}
}
In the GTM, this integral is evaluated by
\aln{
    \langle\mathcal{O}\rangle
    &= \frac{1}{Z}\int_{\Sigma_\tau} dz\,e^{-S(z)}\mathcal{O}(z) \ ,
    \label{observable-deformed-contour}
}
where $\tau$ is the flow time and
the integration contour $\Sigma_\tau$ is defined by solving
the
gradient flow equation \eqref{eq: standard-flow}.


When one applies the HMC algorithm to evaluate this integral,
one has to define a fictitious Hamilton dynamics to update
the configuration on the deformed contour.
In the on-thimble approach,
one defines the Hamilton dynamics on $\Sigma_\tau$,
which is a classical mechanics of a constrained system.
Since the deformed contour $\Sigma_\tau$ is not given explicitly
but has to be determined by solving the flow equation,
complicated procedures are necessary in solving the Hamilton equation
in such a way that the configuration is always on $\Sigma_\tau$.


In the on-axis approach, on the other hand, one rewrites the integral
\eqref{observable-deformed-contour} as
\aln{
  \langle\mathcal{O}\rangle&=
  \frac{1}{Z}\int_{\mathbb{R}^N} dx \,
  \mathrm{det} \,J(x,\tau) \,e^{-S(z(x))}\mathcal{O}(z(x)) \ ,
}
and defines the Hamilton dynamics on the real axis $\mathbb{R}^N$.
Here 
$J(x,\tau)$ is the Jacobi matrix
associated with the change of variables
$x\mapsto z(x,\tau)$,
which obeys the following flow equation
\aln{
\frac{\partial }{\partial \sigma}J_{kl}(x,\sigma)=\overline{H_{km}(z(x,\sigma))}\overline{J_{ml}(x,\sigma)} \ ,
}
where $H_{km}$ is the Hessian defined by \eqref{eq:def-Hessian}.

The fictitious Hamilton dynamics is defined by
the Hamiltonian
\aln{
H=\frac{1}{2}\, \sum_{i=1}^N p_i^2 + \mathrm{Re}\,S(z(x,\tau)) \ ,
\label{eq:H}
}
where $p_i$
are the momentum variables conjugate to
the coordinate variables $x_i$.
Note that there is no constraint that complicates
the Hamilton dynamics unlike the on-thimble approach.
In order to obtain
the expectation value of the observables, one uses the reweighting
\aln{
  \langle\mathcal{O}\rangle=\frac{\langle\mathcal{O}(z(x,\tau)) \, 
    \mathrm{det} \,J(x,\tau) \, e^{-i\, \mathrm{Im} \, S(z(x,\tau))}\rangle_{\rm HMC}}
                 {\langle \mathrm{det} \,J(x,\tau) \, e^{-i \, \mathrm{Im} \, S(z(x,\tau))}\rangle_{\rm HMC}}     \ ,
\label{reweighting}
}
where the expectation value $\langle \ \cdot \ \rangle_{\rm HMC}$
represents an ensemble average over the configurations
generated by the HMC algorithm with the Hamiltonian (\ref{eq:H}).

Note that the force term in the Hamilton equation is given by
\aln{
  F_j&= - \frac{\partial \,\mathrm{Re}S(x,\tau)}{\partial x_j}
  =-\mathrm{Re}\left(f_{i}(x,\tau) J_{ij}(x,\tau)\right) \ ,
\label{eq: force}\\
f_i(x,\tau)
&=\left.\frac{\partial S(z(x,\tau))}{\partial z_i}\right|_{z=z(x,\tau)} \ .
\label{eq: gradient}
}
If one naively uses this formula to calculate the force term,
the Jacobi matrix $J_{ij}(x,\tau)$ that appears in \eqref{eq: force}
requires the computational cost of O($N^2$) for a local system
since it involves matrix-matrix products.
In fact, this can be completely avoided by
using the backpropagation \cite{Fujisawa:2021hxh} as we discuss below.



Let us first note that 
$z(x,\tau)$ and hence $S(z(x,\tau))$ can be regarded
as functions of $z(x,\sigma)$ and $\bar{z}(x,\sigma)$
($0\leq\sigma\leq\tau$)
since $z(x,\sigma+\delta\sigma)\simeq
z(x,\sigma)+\overline{\partial S(z(x,\sigma))}$.
We can therefore define the force at each $\sigma$ as
\aln{
  F_j (\sigma)=\frac{\partial S(z(x,\tau))}{\partial z_j(x,\sigma)} \ ,
  \quad \quad
\bar{F}_j (\sigma)
=\frac{\partial S(z(x,\tau))}{\partial \bar{z}_j(x,\sigma)} \ ,
}
which obeys the relation
\aln{
F_j(\sigma-\delta\sigma)
&=F_i(\sigma)
\frac{\partial z_i(x,\sigma)}{\partial z_j(x, \sigma-\delta\sigma)}
+\bar{F}_i(\sigma)
\frac{\partial \bar{z}_i(x,\sigma)}{\partial z_j(x, \sigma-\delta\sigma)} \ .
\label{eq: backprop}
}
Thus we can calculate $F_j =F_j(0)$ from $f_j = F_j(\tau)$
by solving \eqref{eq: backprop} backward in $\sigma$.
Note that
this is the same procedure as the backpropagation used in machine learning.
Since this procedure
involves the matrix-vector multiplication only,
one can calculate the force term
with the computational cost of
O($N$) for a local system.


\section{Backpropagation with the preconditioner}
\label{sec: forces}

As we have seen in the previous section,
backpropagation \eqref{eq: backprop}
is the key idea that
makes the calculation of the force term \eqref{eq: force} 
in the HMC algorithm feasible in the on-axis approach.
In this section,
we discuss how the backpropagation works with the preconditioned
gradient flow in a way compatible with the idea of the multi-mass solver.

In the case of the preconditioned flow \eqref{eq: precond-flow},
the explicit form of \eqref{eq: backprop} is given by
\aln{
  F_j(\sigma-\delta\sigma)
  &=F_j(\sigma) + \delta\sigma \, G_j(\sigma)  \ , \\
  G_j(\sigma)
  &= 
  F_i(\sigma)
  (\partial_j\mathcal{A}_{ik})\overline{\partial_k S}+\bar{F}_i(\sigma)
  \left[\bar{\mathcal{A}}_{ik}H_{kj}+(\partial_j\bar{\mathcal{A}}_{ik})
    \partial_kS\right] \ .
  \label{eq:backprop-precond}
}
By using the rational approximation 
\eqref{eq:rational_approximation-2} for $\mathcal{A}$,
we can rewrite $G_i(\sigma)$
in \eqref{eq:backprop-precond} as
\aln{
G_j(\sigma)
=& -\sum_{q=1}^Qa_q\left(F^\top(\sigma)
(\bar{H}H+b_q{\bf 1})^{-1}
\bar{H}\right)_{i}(\partial_jH_{ik})\left(
(\bar{H}H+b_q{\bf 1})^{-1}
\overline{\partial S}\right)_{k}
+(\bar{F}^\top \bar{\mathcal{A}}H)_j\nonumber\\
&-\sum_{q=1}^Qa_q\left(\bar{F}^\top (\sigma)
(\bar{H}H+b_q{\bf 1})^{-1}
\right)_{i}(\partial_jH_{ik})\left(\bar{H}
(\bar{H}H+b_q{\bf 1})^{-1}
\partial S\right)_{k} \ . 
}
As in the calculation of the Jacobian using \eqref{del-A-expressions},
the matrix inverse
$(\bar{H}H+b_q{\bf 1})^{-1}$ in the first line
can be applied to
the two vectors $F_j(\sigma)$ and $\overline{\partial _ j S}$
separately, and similarly for the matrix inverse in the second line.
Therefore we can avoid the computational cost of O($Q$) by 
using the idea of the multi-mass solver.

As we discussed in Section \ref{sec:sim-setup},
in order to avoid the ergodicity problem in the GTM,
we have to integrate
over the flow time,
which implies that we have to treat $\tau$
as a dynamical variable in the simulation.
In the HMC algorithm, we therefore have to calculate the force term
for $\tau$ as described in Section 4 of Ref.~\cite{Fujisawa:2021hxh}.
Below we discuss how this can be done efficiently
even in the presence of the preconditioner in the gradient flow.

Since the flow time is discretized as $\tau = N_\tau \delta \tau$
when we solve the gradient flow equation,
the derivative with respect to $\tau$ used in defining the force
$F_\tau(x,\tau)$ for $\tau$ is replaced by the derivative with respect
to the spacing $\delta \tau$.
The force $F_\tau(x,\tau)$ for $\tau$ is defined by
\aln{
  F_\tau(x,\tau)&=
  \mathrm{Re}\Bigl(\partial_jS(x,\tau=N_\tau\delta\tau)
  \, \dot{z}_j(x,\tau=N_\tau\delta\tau) \Bigr) \ , 
\label{eq: tau-force}
}
where $\dot{z}_j(x,\tau=N_\tau\delta\tau)$
is defined by the difference equation
\aln{
   &  (n+1) \, \dot{z}_j(x,(n+1)\, \delta\tau) \nn \\
    =& \  n \,   \dot{z}_j(z,n\, \delta\tau) + \mathcal{A}_{jk}
  \overline{\partial_kS}\nonumber\\
  & + n \, \delta\tau \Bigl[\mathcal{A}_{jk} \, \bar{H}_{ki}
    \, \dot{\bar{z}}_i(x,n\delta\tau) 
    +(\partial_i\mathcal{A}_{jk}) \, 
    \overline{\partial_k S} \, \dot{z}_{i}(x,n\delta\tau)
    +(\bar{\partial}_i\mathcal{A}_{jk}) \, \overline{\partial_k S} \, 
    \dot{\bar{z}}_i(x,n\delta\tau)\Bigr] 
  \label{eq: precond-flow-dotz}
}
with the condition $\dot{z}_j(z,0)=0$.
Note that
the second term on the right-hand side
is
an inhomogeneous term, which does not appear
in the flow equation \eqref{eq:Jacobian_floweq-preconditioned}
for the Jacobi matrix.
Nevertheless, we
can calculate \eqref{eq: tau-force} by backpropagation as follows.

Using a $2N$-dimensional complex vector 
$\zeta^{(n)} =(\dot{z}(x,n\delta\tau) , \dot{\bar{z}}(x,n\delta\tau))^\top$,
\eqref{eq: precond-flow-dotz} can be written formally as
\aln{
\zeta^{(n+1)}=\mathcal{B}^{(n)}\zeta^{(n)}+\beta^{(n)} \ ,
}
where $\mathcal{B}^{(n)}$ and $\beta^{(n)}$ are a $2N\times2N$ matrix
and a $2N$-dimensional vector, respectively, both depending on $\zeta^{(n)}$.
Then, 
the force \eqref{eq: tau-force} for the flow time can be expanded as
\aln{
2F_\tau&=V^{(N_\tau)\top} \zeta^{(N_\tau)}\nonumber\\
&=V^{(N_\tau)\top}(\mathcal{B}^{(N_\tau-1)}\zeta^{(N_\tau-1)}+\beta^{(N_\tau-1)})\nonumber\\
&=V^{(N_\tau)\top}\mathcal{B}^{(N_\tau-1)}(\mathcal{B}^{(N_\tau-2)}\zeta^{(N_\tau-2)}+\beta^{(N_\tau-2)})
+V^{(N_\tau)\top}\beta^{(N_\tau-1)}\nonumber\\
&= V^{(N_\tau)\top} \sum_{n=1}^{N_\tau}
    \left( \prod_{m=0}^{n-1}
    \mathcal{B}^{(m)}\right)  \beta^{(N_\tau-n)} \ ,
    \label{eq:tau-force-2}
}
where
we have introduced a $2N$-dimensional
vector $V^{(N_\tau)}=(v^{(N_\tau)} , \bar{v}^{(N_\tau)})^\top$
with $v_j^{(N_\tau)}=\partial_jS(x,N_\tau\delta\tau)$
and used the initial condition $\zeta^{(0)}=0$.
The product $\prod_{m=0}^{n-1}$ in the last line
should be understood as the time-ordered product, 
in which
$\mathcal{B}^{(m)}$ with smaller $m$ appears on the right.
Here we define a vector
$V^{(n)\top} \equiv V^{(N_\tau)\top}\mathcal{B}^{(N_\tau-1)}\cdots\mathcal{B}^{(n)}$,
which can be obtained recursively by using the relation
\aln{
  V^{(n-1)\top} &= V^{(n)\top} \mathcal{B}^{(n-1)}  \ .
  \label{eq:V-recursion}
}
Then we can obtain the force \eqref{eq:tau-force-2} by
\aln{
  2F_\tau&= \sum_{j=0}^{N_\tau-1} V^{(j+1)\top} \beta^{(j)} \ .
  \label{eq:force-recursion}
  }




More explicitly, the algorithm to calculate the force for $\tau$
can be derived as follows.
Let us represent $V^{(n)}$ as
$V^{(n)}=(v^{(n)} , \bar{v}^{(n)})^\top$.
From \eqref{eq:V-recursion}, we obtain 
\aln{
  v_j^{(n)}&=\frac{n}{n+1}
  \biggl\{v_i^{(n+1)}\Bigl[\delta_{ij}+\delta\tau(\partial_j\mathcal{A}_{ik})
    \overline{\partial_k S}\Bigr]+\bar{v}_i^{(n+1)}
  \delta\tau\Bigl[\bar{\mathcal{A}}_{ik}H_{kj}
    +(\partial_j {\mathcal{A}}_{ik})\partial_kS\Bigr]\biggr\}\ ,
\label{eq: backprop-tau-1}\\
v_j^{(N_\tau)}&=\partial_jS(x,N_\tau\delta\tau) \ .
\label{eq: backprop-tau-3}
}
Then, the force \eqref{eq:force-recursion}
can be obtained as
\aln{
  F_\tau(x,\tau)&=
  \mathrm{Re}\left(f_\tau^{(0)}\right)  \ ,
}
where $f_\tau^{(0)}$ is given by solving
\aln{
f_\tau^{(n)}&=f_\tau^{(n+1)}+\frac{1}{n+1}v_j^{(n+1)}\mathcal{A}_{jk}
\overline{\partial_kS} \ ,
\label{eq: backprop-tau-2}\\
f_\tau^{(N_\tau)}&=0 \ .
}
From \eqref{eq: backprop-tau-1},
we find that the computational cost
of O($Q$) can be avoided by the multi-mass solver here as well.


\section{The parameters of the GTM used in our simulation}
\label{sec: parameters}

In this appendix, we present the parameters chosen for the GTM
such as
the range $[\tau_{\rm min}, \tau_{\rm max}]$ of the flow time
as well as $\tilde{m}$, $m(\tau)$ and $W(\tau)$
that appear in \eqref{eq: Hamiltonian}.

In Table \ref{tab: parameters-1},
we present the choice of
$\tau_{\rm min}$, $\tau_{\rm max}$, $\tilde{m}$
  and $m(\tau)$, which is parameterized as
\begin{align}
m(\tau)&=\exp \left(\sum_{j=0}^{2}a_j\tau^j \right) \ .
\end{align}
The symbols O and P in the ``flow'' column
stand for the original and preconditioned flows, respectively\footnote{For
the preconditioned flow with $N=20$ and $x_{\rm f}=0$, 
the region of $\tau$ is chosen to be
$[0.4,0.8]$ in Fig.~\ref{fig: Jacobian-modulus} for the sake of comparison
with other values of $N$,
whereas it is chosen to be $[0.6,1]$ in Fig.~\ref{fig: final-WF},
which turns out to be more optimal.}.



In Table \ref{tab: parameters-2},
we present
the choice of $W(\tau)$, which is parameterized as
\begin{align}
W(\tau)&=\exp(-15\tau)+\sum_{j=1}^{6}b_j\tau^j \ . 
\end{align}
The first term is
introduced to avoid the dominance of $\tau=0$ that occurs otherwise.
One can see that the functions 
$m(\tau)$ and $W(\tau)$ for the preconditioned flow 
do not have strong dependence on $N$ and $x_{\rm f}$,
which implies that we do not need to fine-tune these functions.

    \begin{table}[H]
      \caption{Parameters for $[\tau_\text{min},\tau_\text{max}]$,
        $\tilde{m}$ and $m(\tau)$}
     \label{tab: parameters-1}
     \centering
      \begin{tabular}{crrrrrrrr}
       \toprule             
       flow &  $N$ & $x_{\rm f}$ & $\tau_\text{min}$ & $\tau_\text{max}$ &
       $\tilde{m}$ & $a_0$ & $a_1$ & $a_2$ \\
       \midrule
       O & $4$ & 0 & 0.02 & 0.2 & 3 & -0.362471 & 9.4008 & 3.33086 \\
                                                      O & 6 & 0  & 0.02 & 0.2 & 3 &  -0.334801 & 17.7419 & -1.32035 \\
                                                      O & 8 & 0 & 0.02 & 0.2 & 3 & -0.600957 & 28.5264 & -13.1596 \\
                                                      O & 10 & 0 & 0.02 & 0.2 & 4 & -0.655057 & 36.2434 & -20.7533 \\
                                                      O & 12 & 0 & 0.02 & 0.2 & 4 & -0.559024 & 42.8481 & -27.0498 \\
                                                      O & 14 & 0 & 0.02 & 0.2 & 4 & -0.5425 & 49.8071 & -36.6114 \\
                                                      O & 16 & 0 & 0.02 & 0.2 & 5 & -0.599934 & 59.8716 & -58.5702 \\
       \midrule
                                                      P & 4 & 0 & 0.4 & 0.8 & 1 & -0.0309437 & 1.42644 & 0.405386 \\
                                                      P  & 6 & 0 & 0.4 & 0.8 & 1 & 0.00490622 & 1.06141 & 0.671549 \\
                                                      P  & 8 & 0 & 0.4 & 0.8 & 1 & 0.0200845 & 0.72403 & 1.00223 \\
                                                      P  & 10 & 0 & 0.4 & 0.8 & 1 & -0.0561224 & 0.930633 & 0.829279 \\
                                                      P  & 12 & 0 & 0.4 & 0.8 & 1 & -0.178908 & 1.27849 & 0.521082 \\
                                                      P  & 14 & 0 & 0.4 & 0.8 & 1 & -0.168445 & 1.24443 & 0.531839 \\
                                                      P  & 16 & 0 & 0.4 & 0.8 & 1 & -0.168445 & 1.24443 & 0.531839 \\
                                                      P  & 18 & 0 & 0.4 & 0.8 & 1 & -0.126564 & 1.03928 & 0.560491 \\
                                                      P  & 20 & 0 & 0.4 & 0.8 & 1 & -0.126564 & 1.03928 & 0.560491 \\
       \midrule                                                      
                                                      P & 20 & -0.8 & 0.4 & 1 & 1 &-0.121107 & 2.53145 & 0.158009 \\
                                   P    & 20  & -0.6 & 0.5 & 1.1 & 1 & -0.0835431 & 2.40247 & 0.228994 \\
                                    P   & 20  & -0.4 & 0.5 & 1.1 & 1 & -0.11638 & 2.56039 & 0.101282 \\
                                    P   & 20  & -0.2 & 0.6 & 1 & 1 & -0.128094 & 2.54973 & 0.118526 \\
                                   P    & 20  & 0 & 0.6 & 1 & 1 & -0.126564 & 1.03928 & 0.560491 \\
                                   P    & 20  & 0.2 & 0.9 & 1.3 & 1 & -0.163594 & 2.69719 & 0.00834929 \\
                                   P    & 20  & 0.4 & 0.9 & 1.3 & 1 & -0.110241 & 2.42494 & 0.211641 \\
                                   P    & 20  & 0.6 & 0.8 & 1.3 & 1 & -0.132109 & 2.56267 & 0.0664781 \\
                                   P    & 20  & 0.8 & 0.8 & 1.3 & 1 & -0.12692 & 2.4658 & 0.165435 \\
       \bottomrule          
      \end{tabular}
    \end{table}
    \begin{table}[H]
     \caption{Parameters for $W(\tau)$}
     \label{tab: parameters-2}
     \centering
      \begin{tabular}{crrrrrrrr}
       \toprule             
          flow &$N$ & $x_{\rm f}$  & $b_1$ & $b_2$ & $b_3$ & $b_4$ & $b_5$ & $b_6$ \\
       \midrule
        O & 4 & 0 & -91.6818& 1328.09& -10070.6 & 42914.7&-94148.5& 82699.2 \\
        O  & 6 & 0 & -73.3069& 934.999& -5039.04 & 15399.9 &-25129.1 & 17155.3\\
                                                     O   & 8 & 0 & -154.67& 2617.17& -17188.6 & 53938.3& -54440.9&-29989.7 \\
                                                      O   & 10 & 0 & -162.532& 3457.74& -27741.1 & 129678& -367262 & 516287 \\
                                                      O   & 12 & 0 & 86.3215& -907.268& 16692.7 &  -116092& 355871 &-402478 \\
                                                      O   & 14 & 0 &1464.1& -29301.3& 325955 &-1942690 & 5900110 &-7172110 \\
                                                      O   & 16 & 0 & 6483.54& -144699 & 1692600 & -10685300 & 34659500 &-45355600 \\
       \midrule
                                                      P & 4 & 0 & -43.3664& 161.375& -352.772 &437.658&-281.151&72.5656 \\
                                                    P  & 6 & 0 & -43.3664& 161.375& -352.772 &437.658&-281.151&72.5656 \\
                                                    P  & 8 & 0 & -43.3664& 161.375&-352.772 &437.658&-281.151&72.5656 \\
                                                    P & 10 & 0 & -43.3664& 161.375&-352.772 &437.658&-281.151&72.5656 \\
                                                    P  & 12 & 0 & -43.3664& 161.375&-352.772  &437.658&-281.151&72.5656 \\
                                                    P  & 14 & 0 & -43.3664& 161.375&-352.772 &437.658&-281.151&72.5656 \\
                                                    P  & 16 & 0 & -43.3664& 161.375&-352.772 &437.658&-281.151&72.5656 \\
                                                    P  & 18 & 0 & -43.3664& 161.375&-352.772 &437.658&-281.151&72.5656 \\
                                                    P  & 20 & 0 & 2143.73& -8632.61& 18071.1 & -20840& 12566.2& -3092.1 \\
    \midrule
                                                    P & 20& -0.8 &-93.4087& 266.479& -399.747 &343.83&-155.337 & 28.5269  \\
                                  P   &   20   & -0.6 & -93.4087& 266.479& -399.747 & 343.83&-155.337 & 28.5269 \\
                                 P  &   20   & -0.4 & -93.7977&266.14& -396.175 & 336.589& -149.815 & 27.0416 \\
                                P  &   20   & -0.2 & -89.7574& 248.492& -358.03 & 294.986&-127.833 &22.5733 \\
                                P  &   20   & 0 & -101.172& 250.908& -360.559 & 300.261 &-132.148 & 23.9199  \\
                                P  &   20   & 0.2 & -96.0689& 294.63& -466.383 & 415.773 &-193.274&36.3518 \\
                                P  &   20   & 0.4 & -105.914& 314.982& -488.542 & 427.174 &-194.103&35.5944 \\
                              P   &   20   & 0.6 & -103.087& 297.241& -447.869 &383.163 &-171.57 &31.2238 \\
                              P  &   20   & 0.8 & -93.3297&  250.673& -347.381 & 273.918 &-113.639 &19.2415 \\
       \bottomrule          
      \end{tabular}
    \end{table}


\newpage

\bibliographystyle{JHEP}
\bibliography{thimble-pcd}

\end{document}